\documentclass[showpacs,preprintnumbers]{revtex4}
\usepackage{amssymb}
\usepackage{amsmath}
\usepackage{graphicx}
\usepackage{dcolumn}
\usepackage{bm}
\usepackage{subfigure}
\usepackage[colorlinks,citecolor=blue, linkcolor=blue,hyperindex,dvipdfm]{hyperref}

\begin{document}

\title{Phase transitions and entropy force of charged de Sitter black holes with cloud of string and  quintessence}
\author{Yubo Ma$^{1}$, Yang Zhang$^{1}$, Ren Zhao$^{1}$, Shuo Cao$^{2*}$, Tonghua Liu$^{2}$, Shubiao Geng $^{2}$, Yuting Liu$^{2}$, Yumei Huang$^{3}$}
\address{ $^{1}$ Institute of Theoretical Physics, Shanxi Datong University, Datong, 037009, China \\$^{2}$ Department of Astronomy, Beijing Normal University, Beijing 100875, China \\$^{3}$ School of Mathematics and Physics, Mianyang Teachers' College, Mianyang 621000, China}

\begin{abstract}
In this paper, we investigate the combined effects of the cloud of strings and quintessence on the thermodynamics of a Reissner-Nordstr\"om-de Sitter black hole. Based on the equivalent thermodynamic quantities considering the correlation between the black hole horizon and the cosmological horizon, we extensively discuss the phase transitions of the space-time. Our analysis prove that similar to the case in AdS space-time, second-order phase transitions could take place under certain conditions, with the absence of first-order phase transition in the charged de Sitter black holes with cloud of string and quintessence. The effects of different thermodynamic quantities on the phase transitions are also quantitatively discussed, which provides a new approach to study the thermodynamic qualities of unstable dS space-time. Focusing on the entropy force generated by the interaction between the black hole horizon and the cosmological horizon, as well as the Lennard-Jones force between two particles, our results demonstrate the strong degeneracy between the entropy force of the two horizons and the ratio of the horizon positions, which follows the surprisingly similar law given the relation between the Lennard-Jones force and the ratio of two particle positions. Therefore, the study of the entropy force between two horizons, is not only beneficial to the deep exploration of the three modes of cosmic evolution, but also helpful to understand the correlation between the microstates of particles in black holes and those in ordinary thermodynamic systems.
\end{abstract}

\maketitle

\section{Introduction}

In the paradigm of de Sitter space-time, the black hole horizon and the cosmological horizon are always considered to be different thermodynamic systems with different thermodynamic quantities and radiation temperature. Therefore, de Sitter space-time does not satisfy the requirements of the stability of thermodynamic equilibrium, which makes it very difficult to investigate the thermodynamics of de Sitter (dS) black holes. Moreover, the discovery of cosmic acceleration is arguably one of the most important developments in modern cosmology, which has also triggered wide spread attention to deeply understand the thermodynamic properties of de Sitter space-time \cite{Mbarek19,Simovic18,Simovic19,Dolan13,Hendi17,Sekiwa06,Kubiznak16,McInerney16,Urano09,Bhattacharya13,
Cai02a,Cai02b,Koberlein94,Azreg-Aiou15,Zhang16a,Zhang14,Zhao14,Li17,Zhang16b,Liu19,Ma18}. In the early 1980's the inflationary universe scenario (Guth 1981), with its predictions of a spatially flat universe and almost scale-invariant density perturbations, changed the cosmological landscape and helped set the stage for the discovery of cosmic acceleration. In the early inflation period, our universe behaved as a quasi-de Sitter space-time. On the other hand, with the inclusion of a mysterious component with negative pressure as a new cosmological component, a large number of dark energy models have been proposed to explain the cosmic acceleration. The simplest candidate for dark energy is the cosmological constant (or vacuum energy density), in which our universe will naturally evolve into a new de Sitter phase with the equation of state (EoS) of dark energy equal to $-1$. This model, the so-called LCDM, has provided excellent agreement with a wide range of astronomical data so far. Finally, there has also been flourishing interest in the duality relation of de Sitter space-time, promoted by the recent success of AdS/CFT correspondence in theoretical physics. Therefore, from observational and theoretical point of view, it is rewarding to obtain a better understanding of the classical and quantum properties of de Sitter space-time.

Now it is widely recognized that in de Sitter space-time, the thermodynamic quantities of the black hole horizon and the cosmological horizon are of a function of the mass $M$, the charge $Q$, and cosmological constant $\Lambda$. Therefore, it is necessary to investigate the thermodynamic properties of thermodynamic quantities, taking into account the correlation between the black hole¡¯s horizon and the cosmological horizon. In the recent analysis of Reissner-Nordstr\"om-de Sitte (RN-dS) space-time \cite{Zhang16b}, the mass $M$, the charge $Q$, and cosmological constant $\Lambda$, which satisfy the first law of thermodynamics, were taken as the function of the state parameters of thermodynamic systems. Based on the above assumptions, one could obtain the equivalent thermodynamic quantities of RN-dS space-time and the corresponding total entropy as

\begin{equation}  \label{1.1}
S=\pi r_c^2[1 + {x^2} + f(x)]
\end{equation}

where $x = {r_+ }/{r_c}$. ${r_+}$ and ${r_c}$ respectively denote the radius of black hole horizon and cosmological horizon

\begin{equation}  \label{1.2}
f(x) = \frac{8}{5}{\left( {1 - {x^3}} \right)^{2/3}} - \frac{{2\left( {4 - 5{x^3} - {x^5}} \right)}}{{5\left( {1 - {x^3}} \right)}}
\end{equation}

On the other hand, the recent observations showed that the dark energy is prevailing in our universe \cite{Komatsu11,Riess04}. The state of equation of dark energy approaches to cosmological constant or vacuum energy \cite{Peebles03}, however, the dynamical dark energy is also possible \cite{Copeland06}, and the quintessence could affect the black hole space-time \cite{Stuchlk05}. Kiselev obtained the modified black hole metric for the Schwarzschild black hole in in quintessence \cite{Kiselev03}. Toshmatov and Xu obtained the solution of  the rotational quintessence black hole and Kerr-Newman black hole \cite{Toshmatov17,Xu17}. The thermodynamics and phase transition of the Schwarzschild black hole surrounded by quintessence matter have been discussed by Tharanath and Ghaderi \cite{Tharanath14,Ghaderi16,Penner16}. Penner has studied the Rerssner-Nordstr\"om black hole situation \cite{Penner16,Thomas12,Shao18}. Li investigate the effects of dark energy on $P-V$ criticality of charged AdS black holes by considering the case of the RN-AdS black holes surrounded by quintessence \cite{Li14}. Employing the effective thermodynamic quantities, the thermodynamic stability of black holes surrounded by quintessence is studied in \cite{Ma17}. Toledo and Chatterjee \cite{Toledo19,Chatterjee19} analyzed the thermodynamics of a Rerssner-Nordstr\"om AdS black hole surrounded cloud of string and quintessence, by calculating different thermodynamic variables, determining their critical values as well as the efficiency of a heat engine. Recent progress has been made in Ref.~\cite{Liu19}, which derived the equivalent thermodynamic quantities of Rerssner-Nordstr\"om dS black hole surrounded by quintessence, and furthermore discussed the thermodynamic properties of this space-time possessing both black hole¡¯s horizon and cosmological horizon.

In this paper, we will focus on the Rerssner-Nordstr\"om dS black hole surrounded cloud of string and quintessence (RN-dSSQ), and obtain its equivalent thermodynamic quantities. The analysis of the heat capacity, constant pressure expansion coefficient, isothermal compression system and Gibbs function show that, second-order phase transitions could take place under certain conditions (similar to the case in AdS space-time), while first-order phase transitions do not exist in RN-dSSQ. More importantly, different from the previous result that the isothermal capacity in spherically symmetric AdS black holes is equal to zero, our findings demonstrates the nonzero heat capacity in the case of RN-dSSQ space-time. Such conclusion is well consistent with the results in the case of general thermodynamic systems. Meanwhile, concentrated on the entropy force generated by the interaction between the black hole horizon and the cosmological horizon, we find it is strongly dependent on the ratio of the two horizons¡¯ positions, following the exact law given the relation between the Lennard-Jones force and the ratio of two particle positions \cite{Johnston00}. This paper is arranged as follows: in Section 2, we briefly introduce the thermodynamic properties of RN-dSSQ space-time, and the condition satisfied by the charge Q and the cosmological constant when the temperature of the two horizons are equal to each other. In Section 3, the equivalent thermodynamic quantities of the black hole horizon and the cosmological horizon in RN-dSSQ space-time are derived, based on which the phase transitions of such pace-time is studied in detail. In Section 4, we discuss the entropy force generated by the interaction between the horizons, as well as its comparison with the Lennard-Jones force between two particles. Finally, we present the main conclusions and discussion in Section 5. The units of $G = \hbar  = {k_B} = c = 1$ are used throughout this work.

\section{The space-time structure}

\label{sec:method}

Let¡¯s start from the well-known Einstein-Maxwell field equation

\begin{equation}  \label{2.1}
{G_{\mu \nu }} + \Lambda {g_{\mu \nu }} = 8\pi G{T_{\mu \nu }}
\end{equation}
with the solution proposed by \cite{Kiselev03} in the presence of a cosmological constant $\Lambda$. The additivity and linearity conditions on the quintessence dark energy stress tensor imply

\begin{equation} \label{2.2}
T_t^t = T_r^r = {\rho _q} + \frac{a}{{{r^2}}}
\end{equation}%

\begin{equation} \label{2.3}
T_\theta ^\theta  = T_\varphi ^\varphi  =  - \frac{1}{2}{\rho _q}(3\omega  + 1)
\end{equation}
where
\begin{equation}  \label{2.4}
{\rho _q} =  - \frac{{3\alpha \omega }}{{2{r^{3(\omega  + 1)}}}}
\end{equation}
and ${\rho _q}$ is the density of quintessence field,  $\omega $ is the state parameter and  $\alpha $ is the normalization parameter related to the density of quintessence field, $a$ is an integration constant which is related to the presence of the cloud of string. The static spherically symmetric solution of Einstein equation for a Reissner-Nordstr\"om-dS black hole in quintessence matters have been obtained by Kiselev as \cite{Kiselev03,Toledo19,Chatterjee19}

\begin{equation}  \label{2.5}
d{s^2} =  - f(r)d{t^2} + {f^{ - 1}}d{r^2} + {r^2}d\Omega _2^2
\end{equation}

With the horizon function

\begin{equation}  \label{2.6}
\begin{aligned}
f(r) = 1 - a - \frac{{2M}}{r} + \frac{{{Q^2}}}{{{r^2}}} - \frac{{{r^2}}}{{{l^2}}} - \alpha {r^{ - 3\omega  - 1}} \end{aligned}
\end{equation}
$M$ and $Q$  are the black hole mass and charge.  $l$ is the curvature radius of dS pace. The parameter $a$  and $\alpha $ codify the presence of the cloud of string and the quintessencer, respectively.

If we set $a = \Lambda  = Q = 0$  and put $\omega  =  - {\textstyle{2 \over 3}}$ the space-time reduces to the Schwarzchild solution surrounded by quintessence whose geodesics were studied in \cite{Uniyal15}. Null trajectories of charged black holes surrounded by quintessence were carried out in \cite{Fernando15}, thermodynamic properties and the Joule-Thomson effect was investigated in \cite{Pradhan17}. The Shadow of rotating charged black holes with quintessence was considered in \cite{Fernando15}.

Now, let us analyze the existence of horizons which are determined by imposing that
\begin{equation}  \label{2.7}
\begin{aligned}
f(r) = 1 - a - \frac{{2M}}{r} + \frac{{{Q^2}}}{{{r^2}}} - \frac{{{r^2}}}{{{l^2}}} - \alpha {r^{ - 3\omega  - 1}} = 0 \end{aligned}
\end{equation}

The black hole has three positive horizons: the black hole event horizon $r_{+}$, the internal (Cauchy) horizon $r_{-}$, and a quintessential cosmological horizon $r_{c}$ \cite{Toledo19}. From Eq.(\ref{2.7}), one may straightforwardly obtain
\begin{equation}  \label{2.8}
\begin{aligned}
M &= (1 - a)\frac{{{r_ + } + {r_c}}}{4}+ \frac{{{Q^2}}}{4}\left( {\frac{1}{{{r_ + }}} + \frac{1}{{{r_c}}}} \right) - \frac{1}{{4{l^2}}}\left( {r_ + ^3 + r_c^3} \right) - \frac{\alpha }{4}\left( {r_ + ^{ - 3\omega } + r_c^{ - 3\omega }} \right)\\
&= (1 - a)\frac{{{r_c}(1 + x)}}{4} + \frac{{{Q^2}(1 + x)}}{{4{r_c}x}} - \frac{{r_c^3}}{{4{l^2}}}\left( {1 + {x^3}} \right) - \frac{\alpha }{4}r_c^{ - 3\omega }\left( {1 + {x^{ - 3\omega }}} \right)\\
& = (1 - a)\frac{{{r_c}x(1 + x)}}{{2(1 + x + {x^2})}} + \frac{{{Q^2}(1 + x + {x^2} + {x^3})}}{{2{r_c}x(1 + x + {x^2})}} + \frac{{\alpha r_c^{ - 3\omega }{x^3}}}{{2(1 - {x^3})}}(1 - {x^{ - 3 - 3\omega }})
\end{aligned}
\end{equation}

\begin{equation}  \label{2.9}
\begin{aligned}
\frac{1}{{{l^2}}} &= \frac{{(1 - a)({r_c} - {r_ + })}}{{\left( {r_c^3 - r_ + ^3} \right)}} - \frac{{{Q^2}({r_c} - {r_ + })}}{{\left( {r_c^3 - r_ + ^3} \right){r_c}{r_ + }}} - \frac{{\alpha \left( {r_c^{ - 3\omega } - r_ + ^{ - 3\omega }} \right)}}{{\left( {r_c^3 - r_ + ^3} \right)}}\\
&= \frac{{(1 - a)}}{{r_c^2(1 + x + {x^2})}} - \frac{{{Q^2}}}{{r_c^4x(1 + x + {x^2})}} - \frac{{\alpha (1 - {x^{ - 3\omega }})}}{{r_c^{3 + 3\omega }(1 - {x^3})}}
\end{aligned}
\end{equation}

Some thermodynamic quantities associated with the cosmological horizon are

\begin{eqnarray} \label{2.10}
S_{c}=\pi r_{c^2},\quad \Phi_{c}=\frac{Q}{r_{c}},\quad T_{c}=- \frac{f^{'}r_{c}}{4\pi }
\end{eqnarray}
where $T_{c}$, $S_{c}$ and $\Phi _{c}$ denote the Hawking temperature, the entropy and the electric potential, respectively.

For the black hole horizon, associated thermodynamic quantities are
\begin{equation}  \label{2.11}
\begin{aligned}
{S_ + } = \pi r_ + ^2,\quad {\Phi _ + } = \frac{Q}{{{r_ + }}},\quad {T_ + } = \frac{{f'({r_ + })}}{{4\pi }} = \frac{1}{{4\pi {r_ + }}}\left( {1 - a - \frac{{{Q^2}}}{{r_ + ^2}} - \frac{{3r_ + ^2}}{{{l^2}}} + 3\omega \alpha r_ + ^{ - 3\omega  - 1}} \right)
\end{aligned}
\end{equation}

When the temperature of the black hole horizon is equal to that of the cosmological horizon, from Eqs. (\ref{2.10})-(\ref{2.11}) the cosmological constant ${l^2}$ satisfies the following expression
\begin{equation}  \label{2.12}
\begin{aligned}
\frac{3}{{{l^2}}} = \frac{{1 - a}}{{r_c^2x}} - \frac{{{Q^2}}}{{r_c^4{x^3}}}\frac{{(1 + {x^3})}}{{(1 + x)}} + 3\omega \alpha \frac{{(1 + {x^{ - 3\omega  - 2}})}}{{r_c^{3\omega  + 3}(1 + x)}}
\end{aligned}
\end{equation}

while from Eqs.(\ref{2.9}) and (\ref{2.12}), the electric potential ${\Phi _c}$ reads \cite{Mellor89}

\begin{equation}  \label{2.13}
\begin{aligned}
\frac{{{Q^2}}}{{r_c^2}} = \frac{{{x^2}(1 - a)}}{{{{(1 + x)}^2}}} + \frac{{3\alpha {x^3}(1 - {x^{ - 3\omega }})}}{{r_c^{1 + 3\omega }{{(1 - x)}^3}{{(1 + x)}^2}}} + 3\omega \alpha \frac{{{x^3}(1 + {x^{ - 3\omega  - 2}})(1 + x + {x^2})}}{{r_c^{3\omega  + 1}(1 + x){{(1 - {x^2})}^2}}} \end{aligned}
\end{equation}

Now the combination of Eqs.(\ref{2.10})-(\ref{2.13}) will generate the radiation temperature of the two horizons
\begin{equation}  \label{2.14}
\begin{aligned}
T=&T_{+}= T_{c}= \frac{1}{{4\pi {r_c}}}\left\{ {2(1 - a)\frac{{1 - x}}{{{{(1 + x)}^2}}}} \right. - 3\alpha r_c^{ - 3\omega  - 1}\frac{{(1 - {x^{ - 3\omega }})(1 + {x^2})}}{{{{(1 - x)}^2}{{(1 + x)}^2}}}\\&- 3\omega \alpha r_c^{ - 3\omega  - 1}\frac{{x(1 - {x^{ - 3\omega  - 3}})}}{{(1 + x)}} - \left. {3\omega \alpha \frac{{(1 + {x^{ - 3\omega  - 2}})(1 - {x^3})(1 + {x^2})}}{{r_c^{3\omega  + 1}(1 + x){{(1 - {x^2})}^2}}}} \right\}
\end{aligned}
\end{equation}

\section{Equivalent thermodynamic quantities}

\label{sec:result}
Considering the fact that interaction between the black hole horizon and the cosmological horizon, the equivalent thermodynamic quantities in the two horizons respectively satisfy the corresponding first law of thermodynamics. More specifically, following the recent study of the thermodynamic properties of dS space-time, the thermodynamic volume can be defined as \cite{Zhang14,Zhao14,Li17,Zhang16b}
\begin{equation}  \label{3.1}
\begin{aligned}
V = \frac{{4\pi }}{3}\left( {r_c^3 - r_ + ^3} \right) = \frac{{4\pi }}{3}r_c^3\left( {1 - {x^3}} \right)
\end{aligned}
\end{equation}

Now, taking the mass $M$, the charge $Q$, and the cosmological constant as the state parameters in the space-time, the first law of thermodynamics can be written as
\begin{equation}\label{3.2}
dM = {T_{eff}}dS - {P_{eff}}dV + {\phi _{eff}}dQ
\end{equation}

Note that the total entropy $S$ is an explicit function of the horizon position, with the definition of
\begin{equation}\label{3.3}
S = \pi r_c^2[1 + {x^2} + f(x)]
\end{equation}

Here the undefined function $f(x)$ represents the extra contribution from the correlations of the two horizons. Inserting Eqs. (\ref{2.8}), (\ref{3.1}) and (\ref{3.3}) into Eq. (\ref{3.2}), the effective temperature of space-time is obtained as
\begin{equation}  \label{3.4}
\begin{aligned}
{T_{eff}} = {\left( {\frac{{\partial M}}{{\partial S}}} \right)_{Q,V}} = &\frac{1}{{2\pi {r_c}(1 + x + {x^2})A(x)}}\left\{ {(1 - a)\left[ {(1 + x)(1 + {x^3}) - 2{x^2}} \right]} \right.\\
 &- \frac{{{Q^2}}}{{r_c^2{x^2}}}\left[ {(1 + x + {x^2})(1 + {x^4}) - 2{x^3}} \right] + \frac{{3\alpha r_c^{ - 3\omega  - 1}{x^2}}}{{(1 - x)}}\left. {\left[ {1 - {x^{ - 3\omega }} - \omega ({x^3} - {x^{ - 3 - 3\omega }})} \right]} \right\}
\end{aligned}
\end{equation}
where
\begin{equation}  \label{3.5}
\begin{aligned}
A(x) = \left[ {(1 - {x^3})f'(x) + 2{x^2}f(x) + 2x(1 + x)} \right]
\end{aligned}
\end{equation}

In the lukewarm case, when the electric charge $Q$ and cosmological constant ${l^2}$ are characterized by Eqs.(\ref{2.12})-(\ref{2.13}), the effective temperature of the space-time is given by

\begin{equation}  \label{3.6}
{T_{eff}} = T
\end{equation}

Combining Eqs. (\ref{2.13}), (\ref{3.4}) and (\ref{3.6}), one may obtain the following expression
\begin{equation}  \label{3.7}
\begin{aligned}
&\frac{{1 - x}}{{{{(1 + x)}^2}}}\left[{1 - a - 3\alpha r_c^{ - 3\omega  - 1}\frac{{(1 - {x^{ - 3\omega }})(1 + {x^2})}}{{2{{(1 - x)}^3}}} - 3\omega \alpha r_c^{ - 3\omega  - 1}\frac{{(1 + 2x + 3{x^2})}}{{2(1 + x){{(1 - x)}^2}}} - 3\omega \alpha \frac{{r_c^{ - 3\omega  - 1}{x^{ - 3\omega  - 2}}{x^2}(3 + 2x + {x^2})}}{{2(1 + x){{(1 - x)}^2}}}} \right]\\
 &= \frac{{2x(1 + {x^4})}}{{(1 + x + {x^2}){{(1 + x)}^2}A(x)}}{ {1 - a - \frac{{3\alpha (1 - {x^{ - 3\omega }})(1 + {x^2})}}{{r_c^{1 + 3\omega }2{{(1 - x)}^3}}}} - \frac{{3\alpha \omega (1 + 2x + 3{x^2})}}{{r_c^{1 + 3\omega }2(1 + x){{(1 - x)}^2}}} - {\frac{{3\alpha \omega {x^2}{x^{ - 3\omega  - 2}}(3 + 2x + {x^2})}}{{r_c^{1 + 3\omega }2(1 + x){{(1 - x)}^2}}}}}
\end{aligned}
\end{equation}

Then we can calculate the expression of $A(x)$ from Eq. (\ref{3.7}) and find it to be
\begin{equation}
\begin{aligned}\label{3.8}
A(x) = \frac{{2x(1 + {x^4})}}{{1 - {x^3}}}
\end{aligned}
\end{equation}

Now the combination of Eqs. (\ref{3.5}) and (\ref{3.8}) will generate
\begin{equation}  \label{3.9}
f'(x) + \frac{{2{x^2}}}{{1 - {x^3}}}f(x) = \frac{{2{x^2}(2{x^3} + {x^2} - 1)}}{{{{(1 - {x^3})}^2}}}
\end{equation}

with the corresponding solution as (when taking the boundary condition $f(0) = 0$)
\begin{equation}  \label{3.10}
f(x) = \frac{8}{5}{\left( {1 - {x^3}} \right)^{2/3}} - \frac{{2\left( {4 - 5{x^3} - {x^5}} \right)}}{{5\left( {1 - {x^3}} \right)}}
\end{equation}
One should note that such solution is well consistent with that in the RN-dS space-time \cite{Zhang16b}.

Now we will focus on the derivation of the equivalent thermodynamic quantities in RN-dSSQ space-time. First of all, we get the effective temperature of RN-dSSQ space-time from Eqs. (\ref{3.4}) and (\ref{3.10})
\begin{equation}  \label{3.11}
\begin{aligned}
 {T_{eff}}=  &\frac{{1 - x}}{{4\pi {r_c}x(1 + {x^4})}} \left\{{(1 - a)\left[ {(1 + x)(1 + {x^3}) - 2{x^2}} \right]} - \frac{{{Q^2}}}{{r_c^2{x^2}}}\left[ {(1 + x + {x^2})(1 + {x^4}) - 2{x^3}} \right]\right.\\
 & + \frac{{3\alpha r_c^{ - 3\omega  - 1}{x^2}}}{{(1 - x)}}\left. {\left[ {1 - {x^{ - 3\omega }} - \omega ({x^3} - {x^{ - 3 - 3\omega }})} \right]} \right\}\\
= &\frac{1}{{4\pi {r_c}{x^3}(1 + {x^4})}}\left\{ {(1 - a){x^2}(1 - 3{x^2} + 3{x^3}} \right. - {x^5})- {\phi ^2}(1 - 3{x^3} + 3{x^4} - {x^7})\\
&- \frac{{{\beta _c}{x^4}}}{{2\pi \omega }}\left. {\left[ {1 - {x^{ - 3\omega }} - \omega ({x^3} - {x^{ - 3 - 3\omega }})} \right]} \right\} \\
= &\frac{{B(x,\omega )}}{{4\pi {r_c}{x^3}(1 + {x^4})}}
\end{aligned}
\end{equation}

where
\begin{equation}  \label{3.12}
\begin{aligned}
&\frac{{3\alpha }}{{r_c^{1 + 3\omega }}} = \frac{{3\alpha 4\pi r_c^2}}{{4\pi r_c^{3 + 3\omega }}} =  - \frac{{{A_c}{\rho _c}}}{{2\pi \omega }} =  - \frac{{{\beta _c}}}{{2\pi \omega }},\quad \phi  = \frac{Q}{{{r_c}}},\\
B(x,\omega ) = {x^2}(1 - a)(1 - 3{x^2}& + 3{x^3} - {x^5}) - {\phi ^2}(1 - 3{x^3} + 3{x^4} - {x^7})- \frac{{{\beta _c}{x^4}}}{{2\pi \omega }}\left[ {1 - {x^{ - 3\omega }} - \omega ({x^3} - {x^{ - 3 - 3\omega }})} \right]
\end{aligned}
\end{equation}

In order to clearly see the effect of relevant parameters on the effective temperature, we illustrate an example of the $T_{eff}-x$ diagram with different value of $a$, $\omega$, $\beta _{c}$ and $\phi$, which are explicitly shown in Fig. 1 (by fixing $r_{c} = 1$). It is obvious that, compared with the other two parameters ($\beta _{c}$ and $\omega$), different values of a and $\phi$ (the electric potential at the cosmological horizon) play a more important role in the determination of the maximum effective temperature and its corresponding position $x$. More specifically, when $\phi$ is fixed, the maximum value of the effective temperature of the system will decrease with $a$. Such tendency can also be seen from the behavior of the effective temperature as a function of $\phi$, in term of different $a$.

\begin{figure}[htp]
\centering
\begin{minipage}[t]{0.85\textwidth}
\subfigure[\quad $\phi=0.01, \beta _{c}=0.1, \omega=-1/2$]{\includegraphics[width=0.30\textwidth]{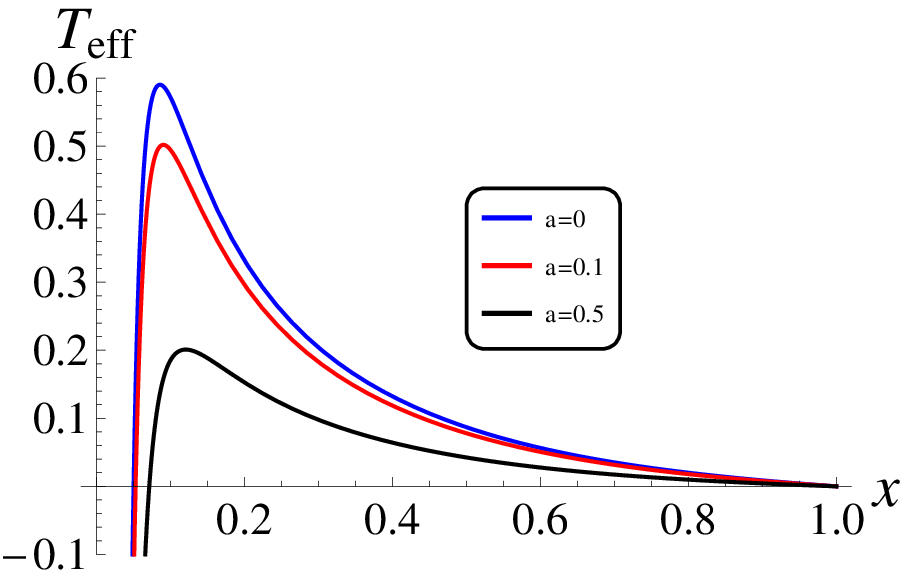}}
\subfigure[\quad$a=0.1, \phi=0.01, \beta _{c}=0.1$]{\includegraphics[width=0.30\textwidth]{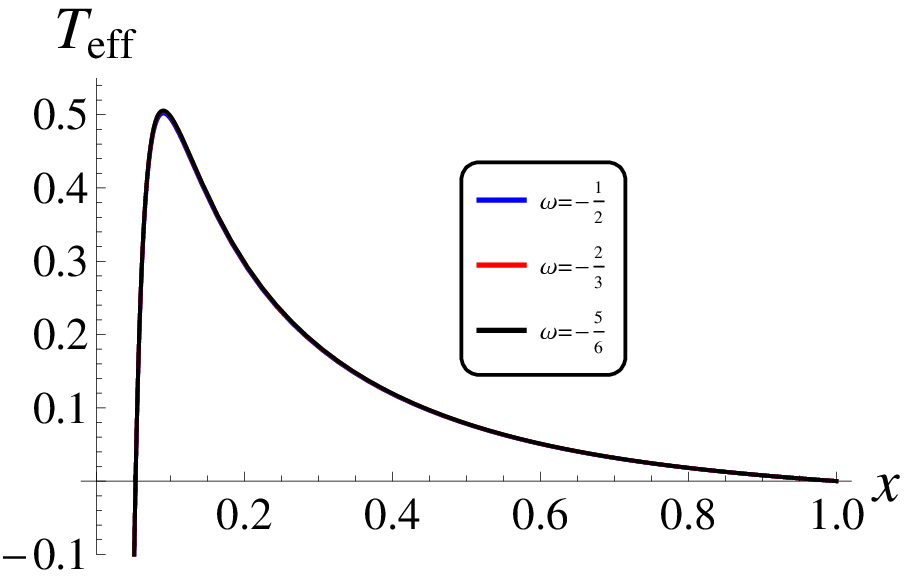}}\newline
\subfigure[\quad$a=0.1, \phi=0.01, \omega=- 1/2$]{\includegraphics[width=0.30\textwidth]{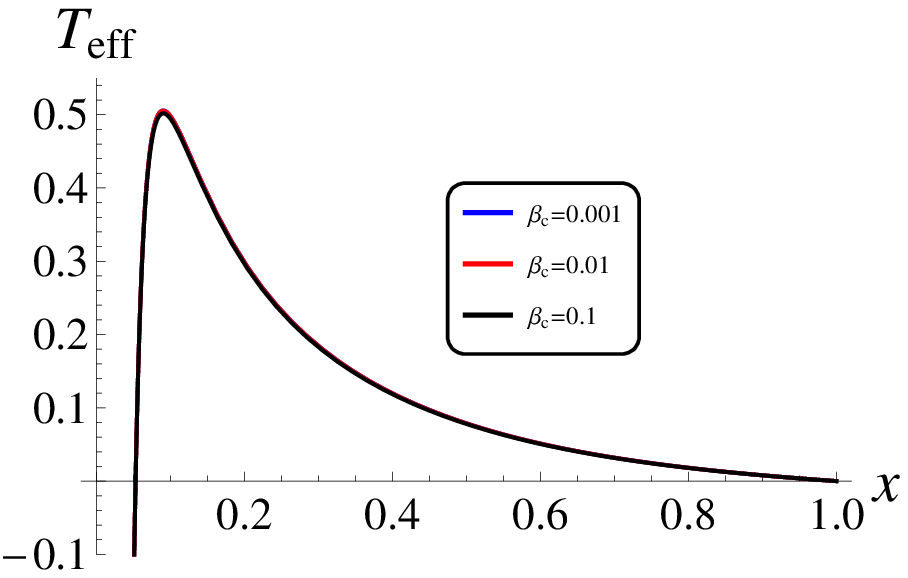}}
\subfigure[\quad$a=0.1, \beta _{c}=0.1, \omega=- 1/2$]{\includegraphics[width=0.29\textwidth]{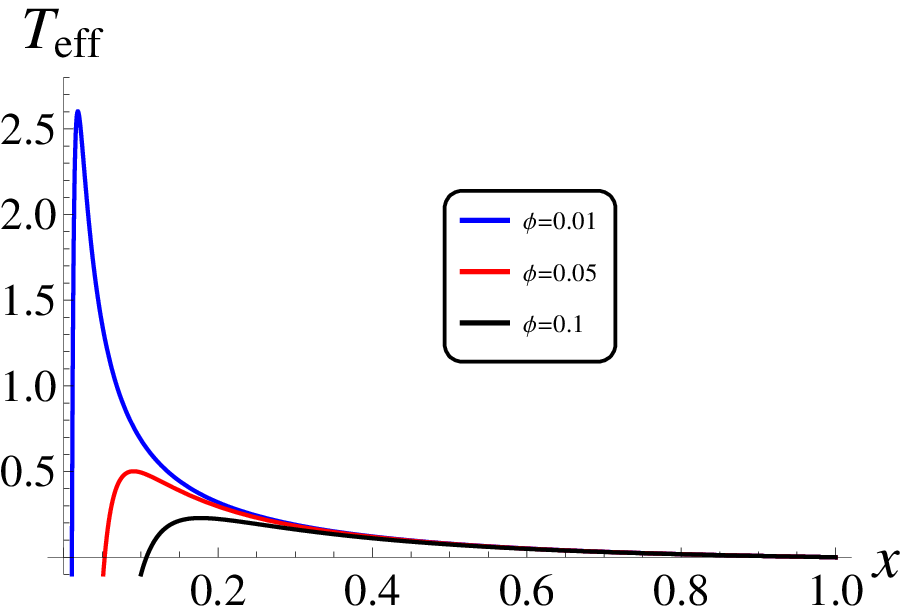}}\newline
\caption{The behavior of the effective temperature as a function of $x$, (a) For different $a$, (b) for different $\Omega$, (c) For different $\beta _{c}$ , and (d) for different $\phi$, when the parameter $r_{c}$ is fixed at 1.}
\label{fig1}
\end{minipage}
\end{figure}

Secondly, the effective pressure of RN-dSSQ space-time can be derived from Eq. (\ref{3.2}), which is given by
\begin{equation}  \label{3.13}
\begin{aligned}
{P_{eff}} =  - {\left( {\frac{{\partial M}}{{\partial V}}} \right)_{Q,S}} = \frac{{(1 - x)D(x,\omega )}}{{8\pi r_c^2x(1 + {x^4})}}
\end{aligned}
\end{equation}
where
\begin{equation}  \label{3.14}
\begin{aligned}
D(x,\omega ) = & - \left[ {(1 - a)x(1 + x) - \frac{{{\phi ^2}(1 + x + {x^2} + {x^3})}}{x} + \frac{{{\beta _c}{x^3}(1 - {x^{3 - 3\omega }})}}{{2\pi (1 - x)}}} \right](x + f'(x)/2)\\
&+ \left[ {\frac{{(1 - a)(1 + 2x)}}{{(1 + x + {x^2})}} - \frac{{{\phi ^2}(1 + 2x + 3{x^2})}}{{{x^2}(1 + x + {x^2})}} - \frac{{{\beta _c}{x^2}(1 - (2 - \omega ){x^{5 - 3\omega }} + (1 - \omega ){x^{8 - 3\omega }})}}{{2\pi \omega (1 - {x^3})(1 - x)}}} \right](1 + {x^2} + f(x))
\end{aligned}
\end{equation}

\begin{figure}[htp]
\centering
\begin{minipage}[t]{0.85\textwidth}
\subfigure[\quad $\phi=0.01, \beta _{c}=0.1, \omega=-1/2$]{\includegraphics[width=0.30\textwidth]{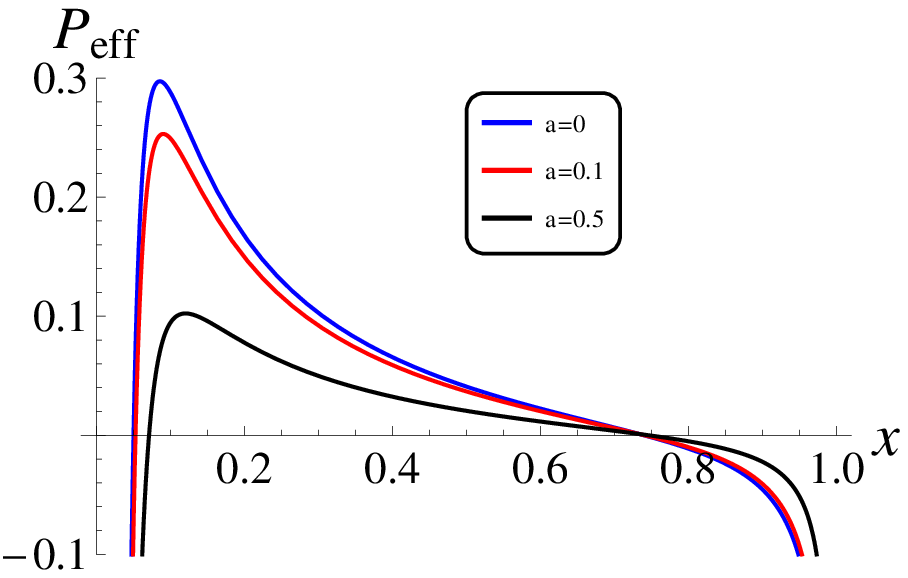}}
\subfigure[\quad$a=0.1, \phi=0.01, \beta _{c}=0.1$]{\includegraphics[width=0.30\textwidth]{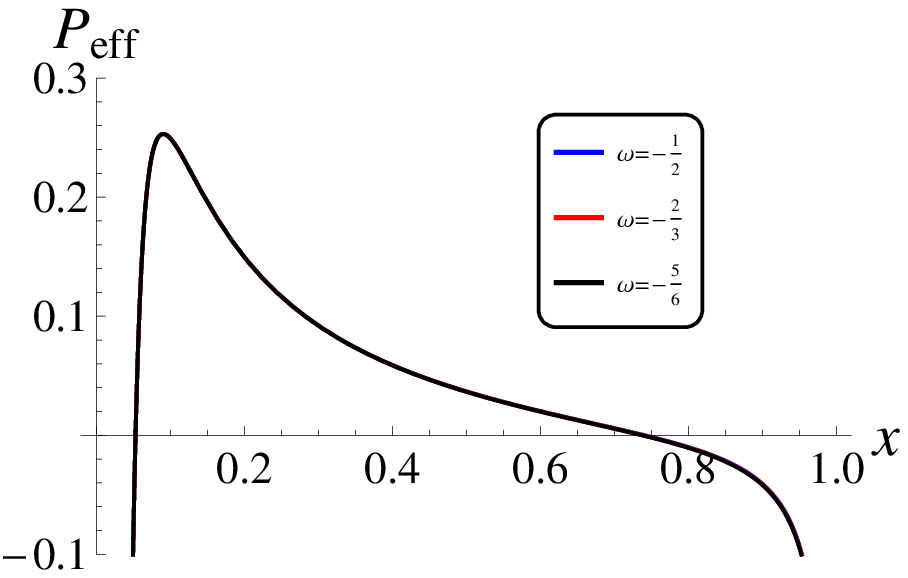}}\newline
\subfigure[\quad$a=0.1, \phi=0.01, \omega=- 1/2$]{\includegraphics[width=0.30\textwidth]{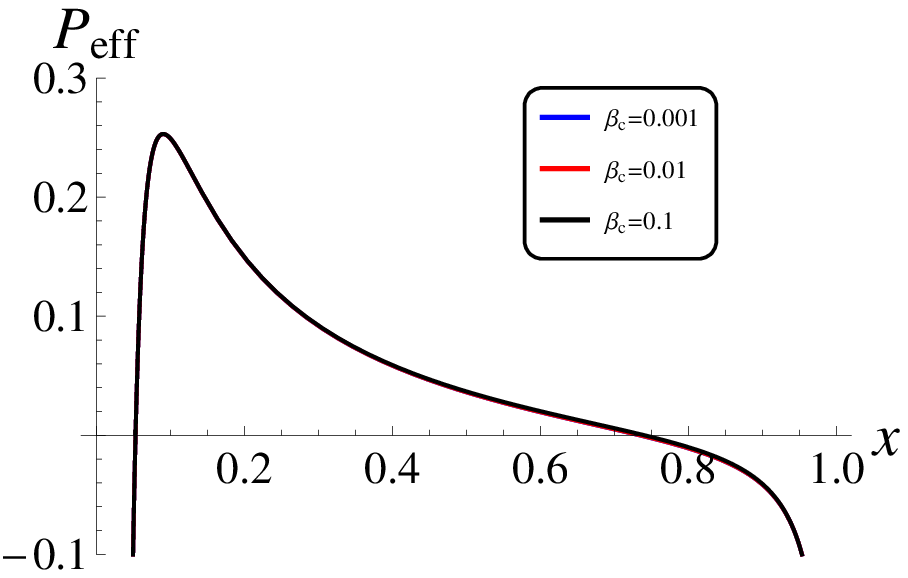}}
\subfigure[\quad$a=0.1, \beta _{c}=0.1, \omega=- 1/2$]{\includegraphics[width=0.29\textwidth]{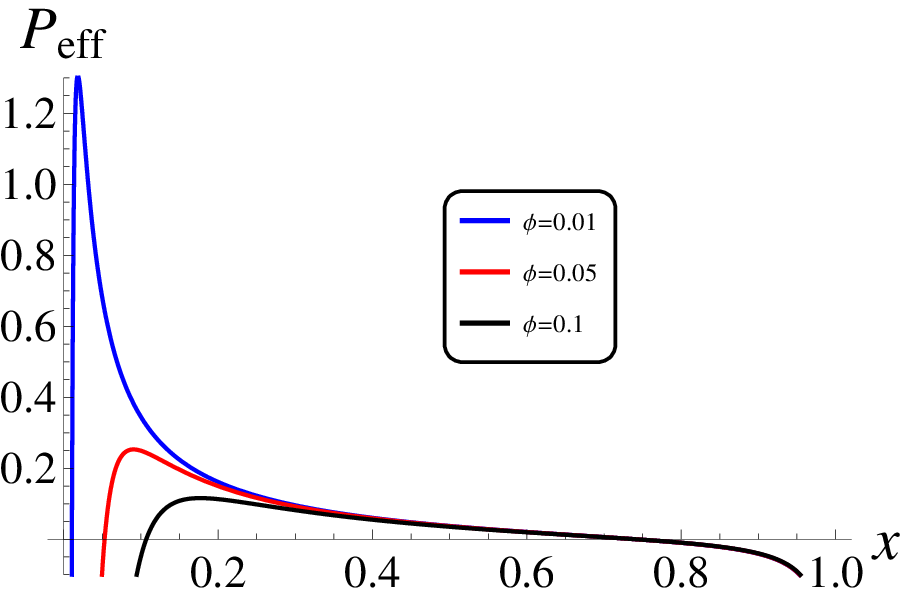}}\newline
\caption{ The same as Fig. 1, but for the behavior of the effective pressure $P_{eff}$ as a function of $x$.}
\label{fig2}
\end{minipage}
\end{figure}

Similarly, we also analyze the behavior of the effective pressure $P_{eff}$. The results are shown in Fig. \ref{fig2}, which quantify the effects of these parameters on the effective pressure of RN-dSSQ space-time. One can clearly see that the maximum value of the effective pressure will significantly will decrease with the value of a and $\phi$, which is quite similar to the behavior of effective temperature illustrated in Fig. \ref{fig1}.

\begin{figure}[htp]
\centering
\begin{minipage}[t]{0.85\textwidth}
\subfigure[\quad $\phi=0.01, \beta _{c}=0.1, \omega=-1/2$]{\includegraphics[width=0.32\textwidth]{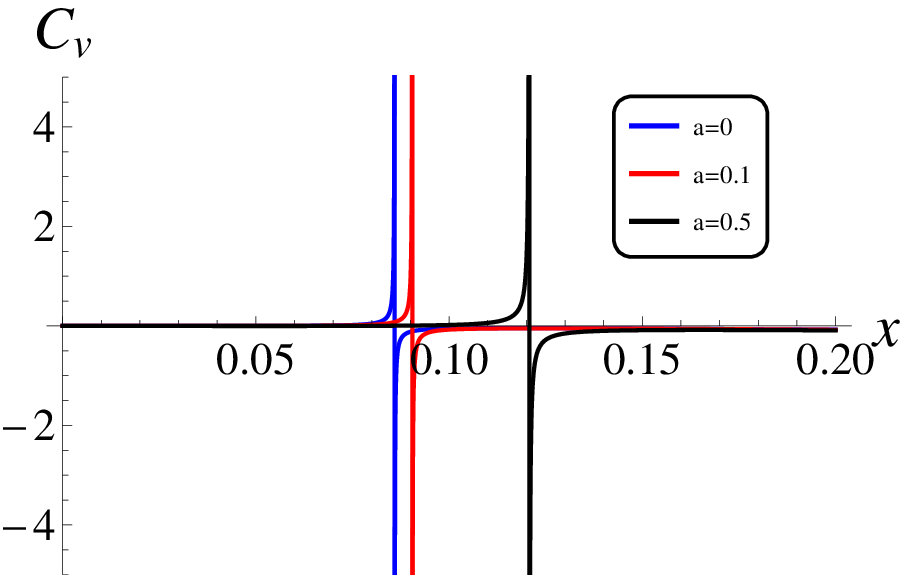}}
\subfigure[\quad$a=0.1, \phi=0.01, \beta _{c}=0.1$]{\includegraphics[width=0.30\textwidth]{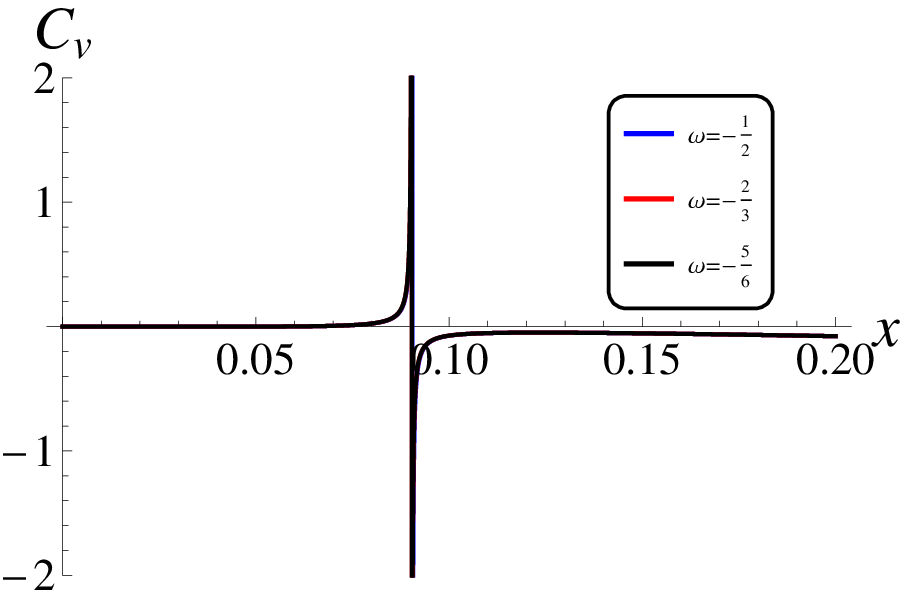}}\newline
\subfigure[\quad$a=0.1, \phi=0.01, \omega=- 1/2$]{\includegraphics[width=0.30\textwidth]{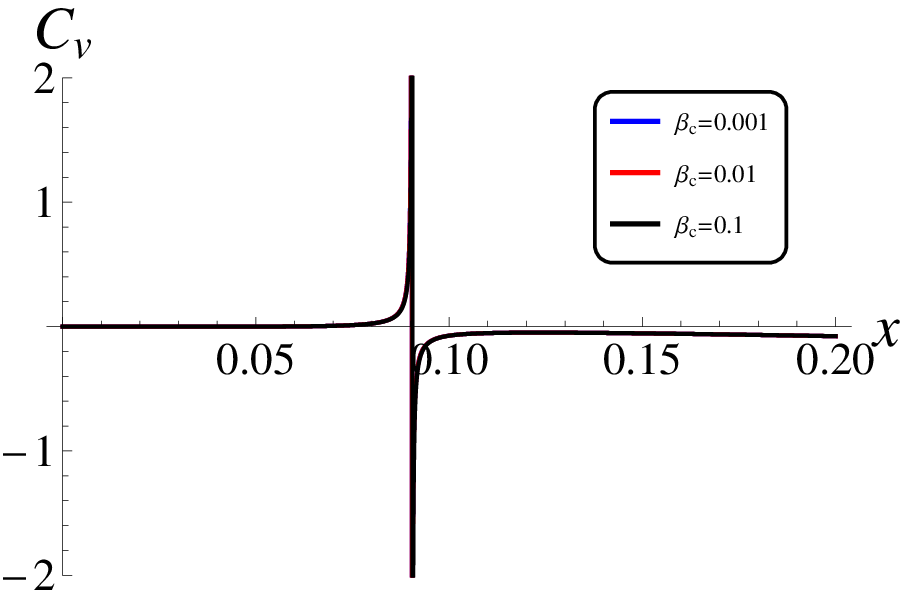}}
\subfigure[\quad$a=0.1, \beta _{c}=0.1, \omega=- 1/2$]{\includegraphics[width=0.30\textwidth]{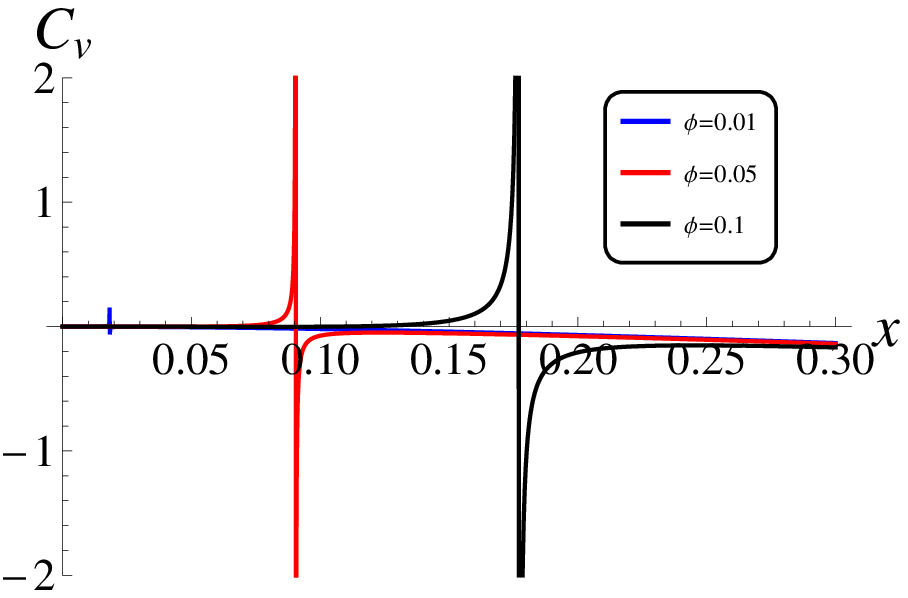}}\newline
\caption{The behavior of the heat capacity $C_{v}$ as a function of $x$ (with $\pi r_c^2 = 1)$}
\label{fig3}
\end{minipage}
\end{figure}

Thirdly, we turn to the behavior of the heat capacity of RN-dSSQ space-time. On the one hand, the heat capacity of isovolumetric systems can be written as

\begin{equation}  \label{3.15}
{C_V} = {T_{eff}}{\left( {\frac{{\partial S}}{{\partial {T_{eff}}}}} \right)_V} = \frac{{2\pi r_c^2B(x,\omega )}}{{1 - {x^3}}}\frac{{{x^2}{{(1 + {x^4})}^2}}}{{B'(x,\omega )x(1 - {x^3})(1 + {x^4}) - B(x,\omega )(3 - 2{x^3} + 7{x^4} - 6{x^7})}}
\end{equation}

with the differential of $B$ given by $B'(x,\omega ) = \frac{{\partial B(x,\omega )}}{{\partial x}}$. When taking $\pi r_c^2 = 1$, one can get the ${C_V} - x$ graphs with different combinations of relevant parameters ($a$, $\phi$, $\beta _{c}$, $\omega$) from the above equations. Comparing the two cases illustrated in Fig. \ref{fig3}, one could easily find that the effective isovolumetric heat capacity of RN-dSSQ space-time is not equal to zero, which is quite different from with the case of ${C_V} = 0$ in AdS space-time. However, such finding in RN-dSSQ space-time agrees very well with the previous result obtained in the framework of ordinary thermodynamic systems, which strongly implies the strong consistency between the two types of thermodynamic systems. On the other hand, another important thermodynamical quantity is the heat capacity at constant pressure, which could be defined as
\begin{equation}  \label{3.16}
\begin{aligned}
{C_{{P_{eff}}}} = {T_{eff}}{\left( {\frac{{\partial S}}{{\partial {T_{eff}}}}} \right)_{{P_{eff}}}} = 2\pi r_c^2\frac{{B(x,\omega )E(x,\omega )}}{{F(x,\omega )}} \end{aligned}
\end{equation}
where
\begin{equation}  \label{3.17}
\begin{aligned}
&E(x,\omega )= \left[ {(1 - x)D'(x,\omega ) - 2\frac{{(1 + 5{x^4} - 4{x^5})D(x,\omega )}}{{x(1 + {x^4})}}} \right][1 + {x^2} + f(x)] + (1 - x)D(x,\omega )(2x + f'(x))\\
&F(x,\omega ) = 2(1 - x)D(x,\omega )\left[ {B'(x,\omega ) - \frac{{B(x,\omega )(3 + 7{x^4})}}{{x(1 + {x^4})}}} \right]- B(x,\omega )\left[ {(1 - x)D'(x,\omega ) - \frac{{(1 + 5{x^4} - 4{x^5})D(x,\omega )}}{{x(1 + {x^4})}}} \right]
\end{aligned}
\end{equation}

\begin{figure}[htp]
\centering
\begin{minipage}[t]{0.85\textwidth}
\subfigure[\quad $\phi=0.01, \beta _{c}=0.1, \omega=-1/2$]{\includegraphics[width=0.30\textwidth]{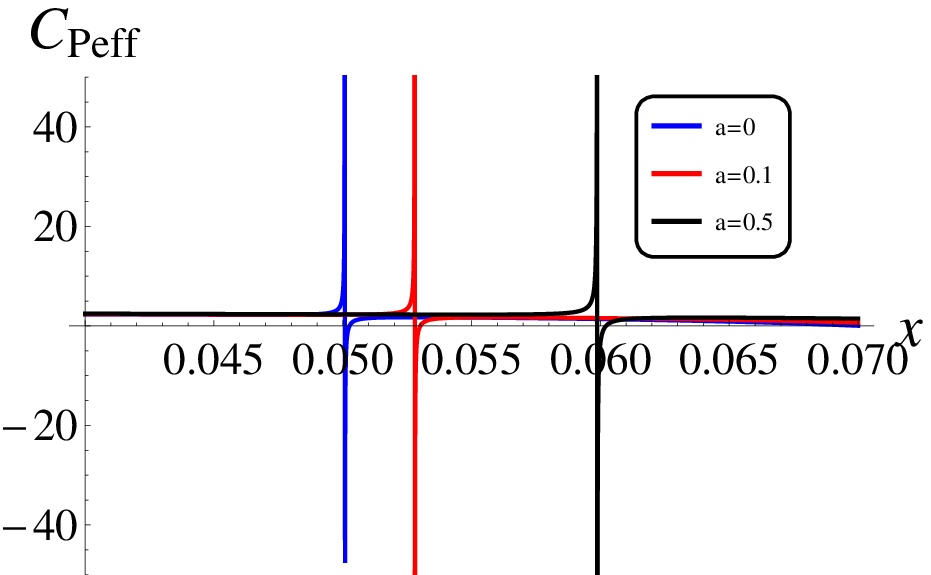}}
\subfigure[\quad$a=0.1, \phi=0.01, \beta _{c}=0.1$]{\includegraphics[width=0.30\textwidth]{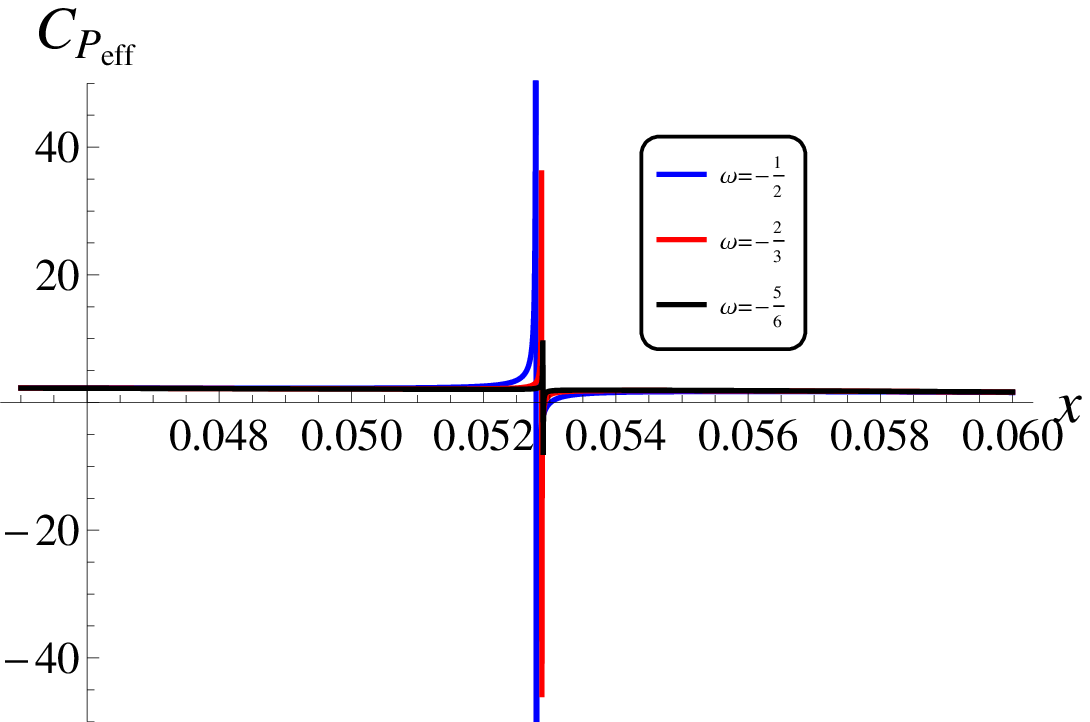}}\newline
\subfigure[\quad$a=0.1, \phi=0.01, \omega=- 1/2$]{\includegraphics[width=0.30\textwidth]{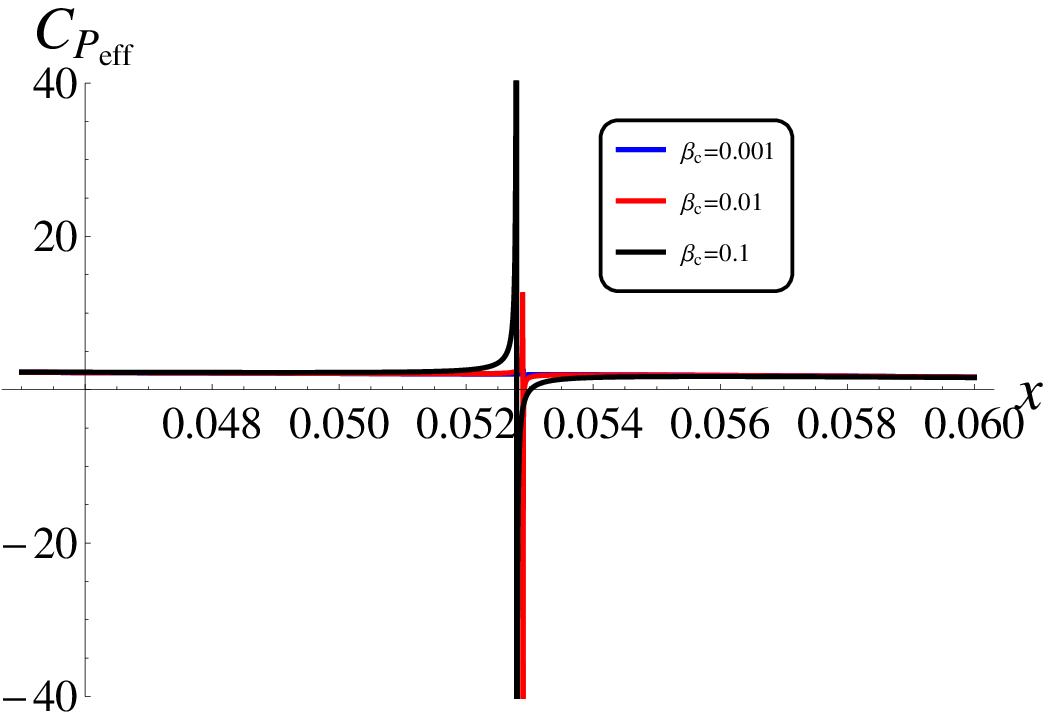}}
\subfigure[\quad$a=0.1, \beta _{c}=0.1, \omega=- 1/2$]{\includegraphics[width=0.30\textwidth]{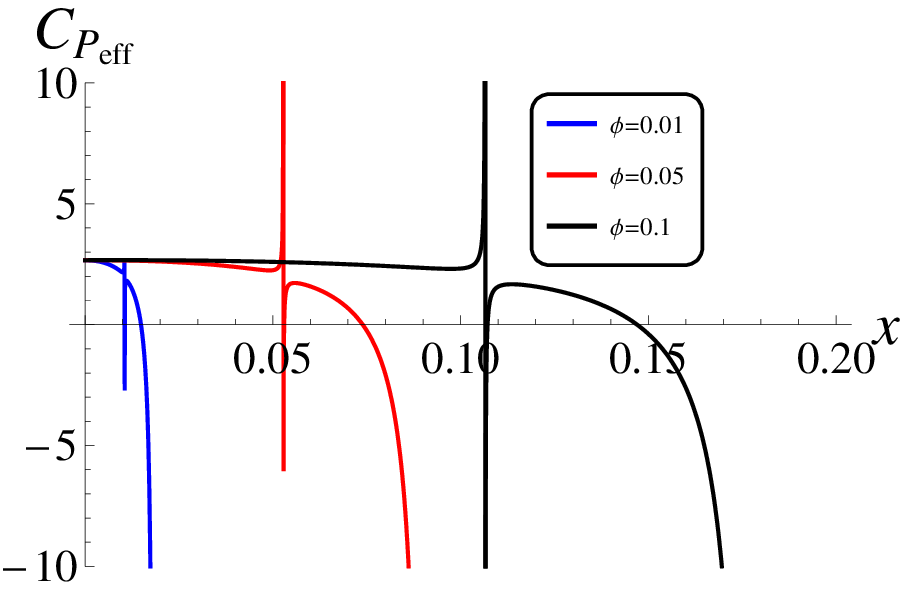}}\newline
\caption{The same as Fig. 3, but for the behavior of the heat capacity $C_{P_{eff}}$ as a function of $x$.}
\label{fig4}
\end{minipage}
\end{figure}

The $C_{V} - x$ graphs with different combinations of relevant parameters ($a$, $\phi$, $\beta _{c}$, $\omega$) are presented in Fig. \ref{fig4}.

\begin{figure}[htp]
\centering
\begin{minipage}[t]{0.85\textwidth}
\subfigure[\quad $\phi=0.01, \beta _{c}=0.1, \omega=-1/2$]{\includegraphics[width=0.30\textwidth]{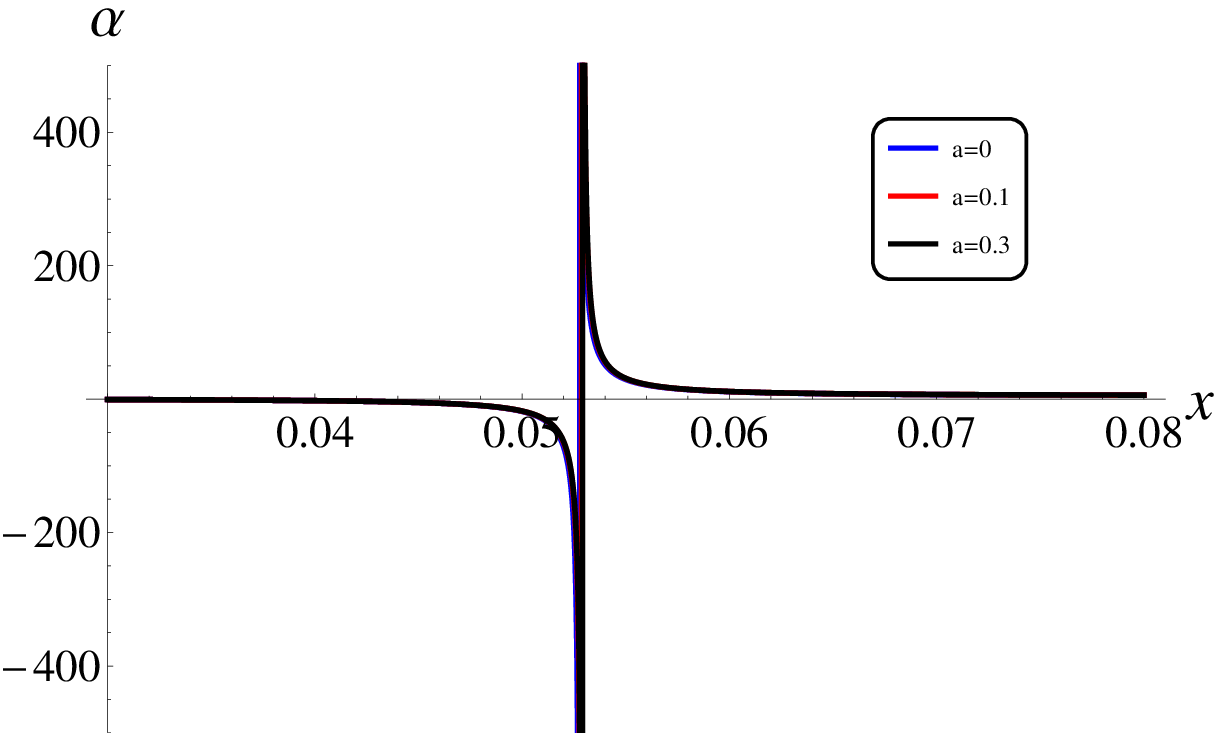}}
\subfigure[\quad$a=0.1, \phi=0.01, \beta _{c}=0.1$]{\includegraphics[width=0.30\textwidth]{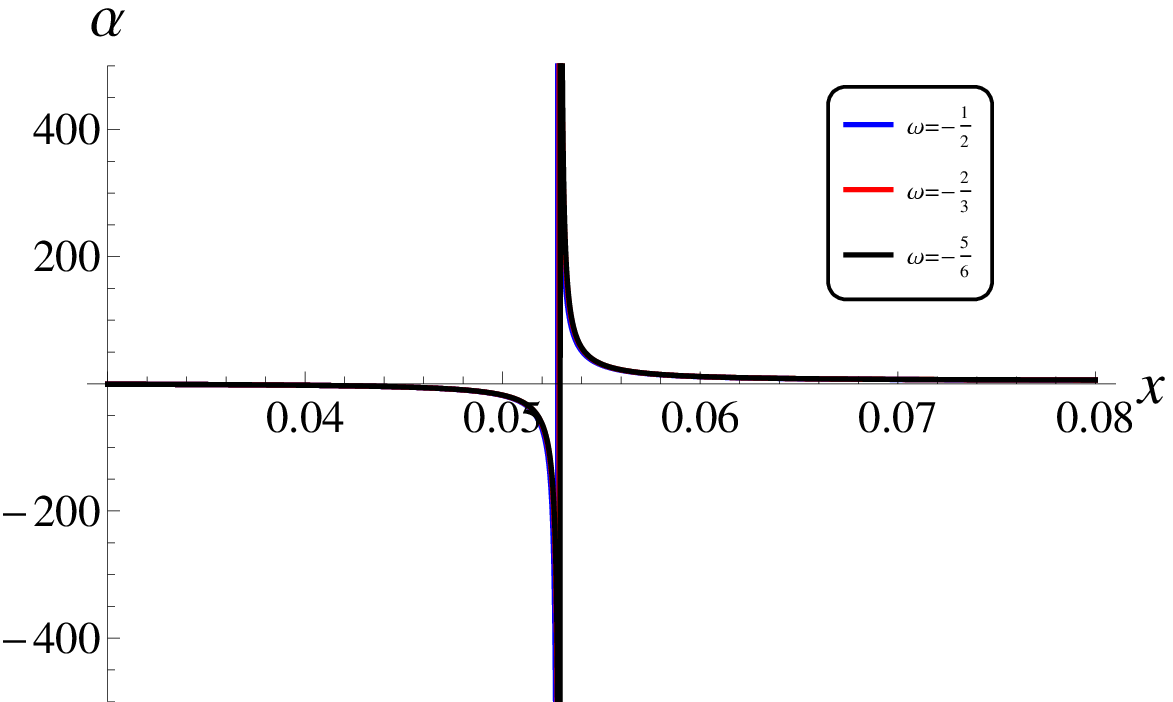}}\newline
\subfigure[\quad$a=0.1, \phi=0.01, \omega=- 1/2$]{\includegraphics[width=0.30\textwidth]{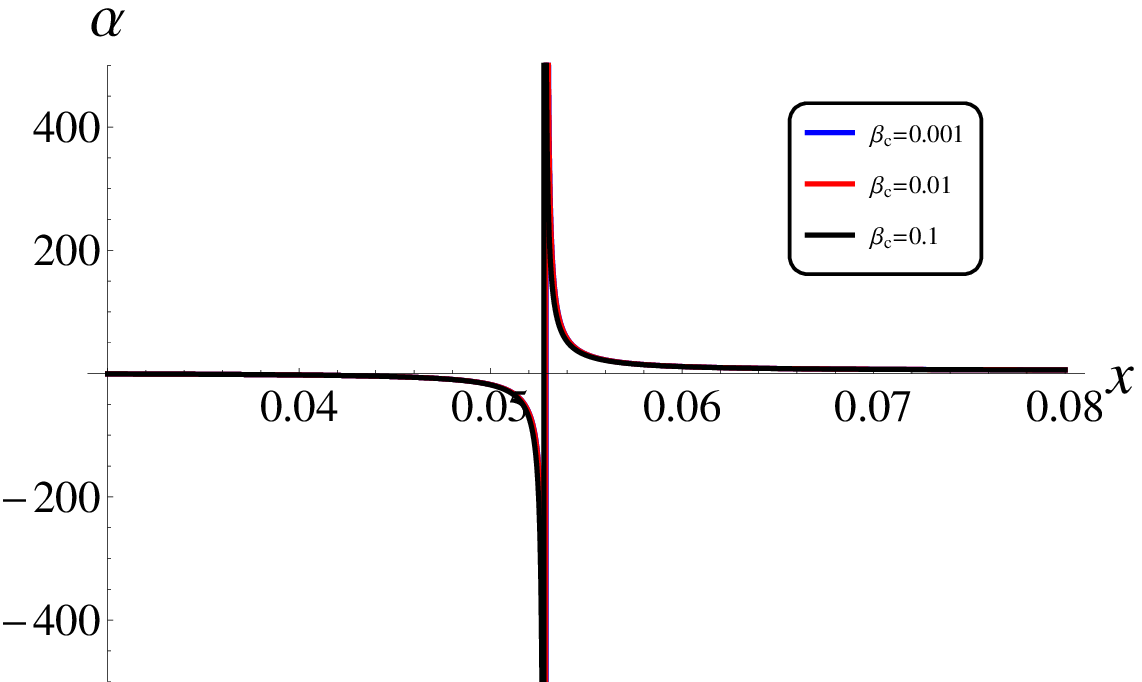}}
\subfigure[\quad$a=0.1, \beta _{c}=0.1, \omega=- 1/2$]{\includegraphics[width=0.30\textwidth]{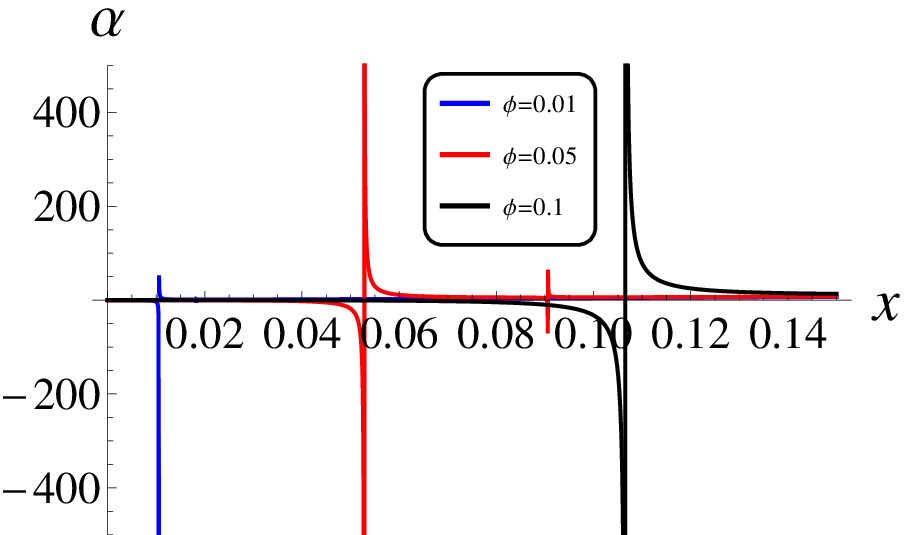}}\newline
\caption{The analog of volume expansion coefficient varying with $x$ ($\pi r_c^2 = 1$ ), based on different combinations of relevant parameters ($a$, $\phi$, $\beta _{c}$, $\omega$).}
\label{fig5}
\end{minipage}
\end{figure}

\begin{figure}[htp]
\centering
\begin{minipage}[t]{0.85\textwidth}
\subfigure[\quad $\phi=0.01, \beta _{c}=0.1, \omega=-1/2$]{\includegraphics[width=0.30\textwidth]{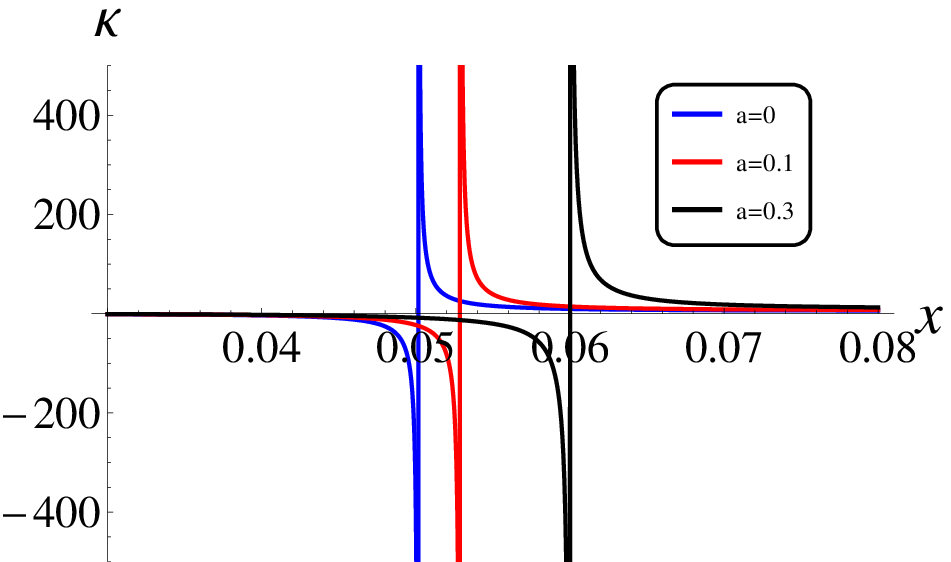}}
\subfigure[\quad$a=0.1, \phi=0.01, \beta _{c}=0.1$]{\includegraphics[width=0.30\textwidth]{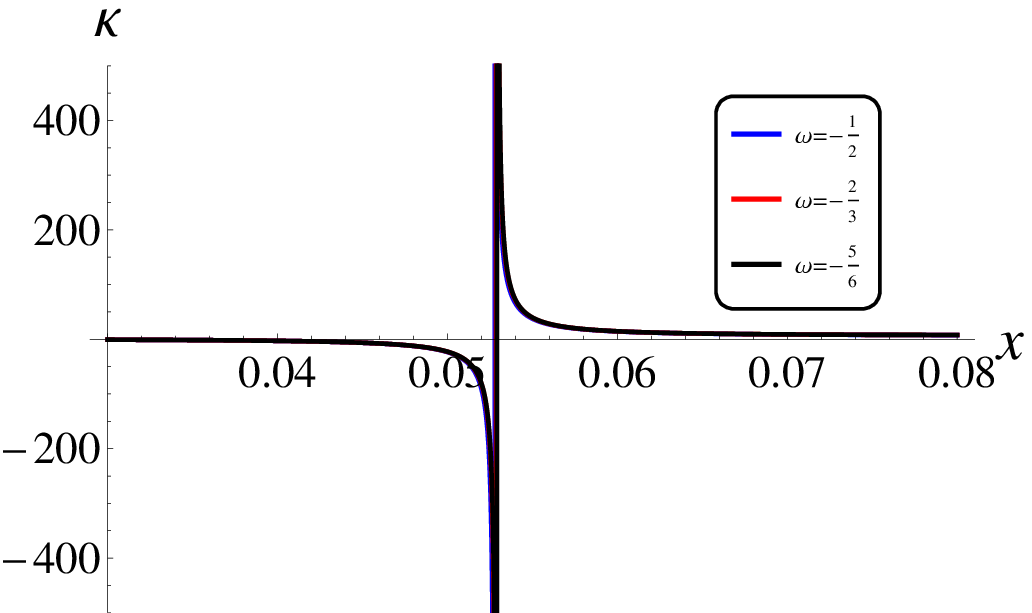}}\newline
\subfigure[\quad$a=0.1, \phi=0.01, \omega=- 1/2$]{\includegraphics[width=0.30\textwidth]{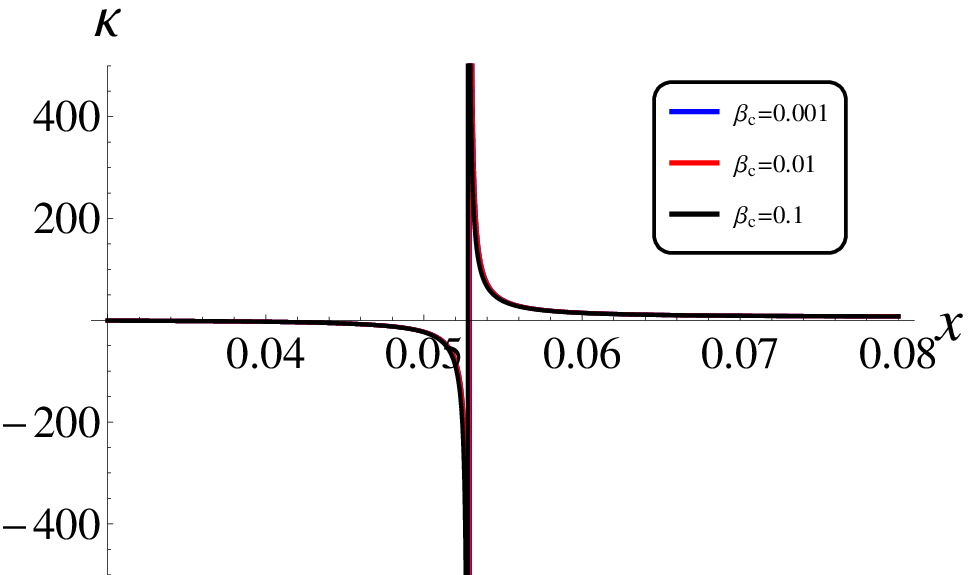}}
\subfigure[\quad$a=0.1, \beta _{c}=0.1, \omega=- 1/2$]{\includegraphics[width=0.30\textwidth]{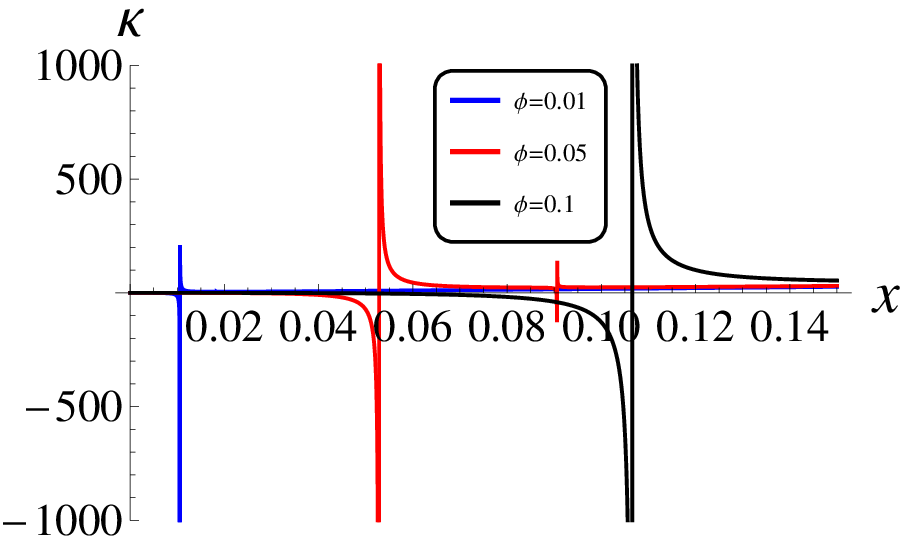}}\newline
\caption{The same as Fig. 5, but for the behavior of the analog of isothermal compressibility.}
\label{fig6}
\end{minipage}
\end{figure}

Now we are at a crossroads in the thermodynamical study of critical phenomena at the critical point, as well as the phase transitions with different space-time parameters. More specifically, in the framework of RN-dSSQ space-time considered in this analysis, the analog of volume expansion coefficient given by

\begin{equation}  \label{3.18}
\begin{aligned}
\alpha & = \frac{1}{V}{\left( {\frac{{\partial V}}{{\partial {T_{eff}}}}} \right)_{{P_{eff}}}} = \frac{1}{V}\frac{{{{\left( {\frac{{\partial V}}{{\partial {r_c}}}} \right)}_x}{{\left( {\frac{{\partial {P_{eff}}}}{{\partial x}}} \right)}_{{r_c}}} - {{\left( {\frac{{\partial V}}{{\partial x}}} \right)}_{{r_c}}}{{\left( {\frac{{\partial {P_{eff}}}}{{\partial {r_c}}}} \right)}_x}}}{{{{\left( {\frac{{\partial {P_{eff}}}}{{\partial x}}} \right)}_{{r_c}}}{{\left( {\frac{{\partial {T_{eff}}}}{{\partial {r_c}}}} \right)}_x} - {{\left( {\frac{{\partial {P_{eff}}}}{{\partial {r_c}}}} \right)}_x}{{\left( {\frac{{\partial {T_{eff}}}}{{\partial x}}} \right)}_{{r_c}}}}}\\
& = \frac{{12\pi {r_c}{x^3}(1 + {x^4})}}{{F(x,\omega )}}\left( {(1 - x)D'(x,\omega ) - \frac{{(1 + 5{x^4} - 4{x^5})D(x,\omega )}}{{x(1 + {x^4})}} - 2\frac{{(1 - x)D(x,\omega )}}{{1 - {x^3}}}{x^2}} \right)
\end{aligned}
\end{equation}

while the analog of isothermal compressibility can be derived as
\begin{equation}  \label{3.19}
\begin{aligned}
{\kappa _{{T_{eff}}}} &=  - \frac{1}{V}{\left( {\frac{{\partial V}}{{\partial {P_{eff}}}}} \right)_{{T_{eff}}}} = \frac{1}{V}\frac{{{{\left( {\frac{{\partial V}}{{\partial {r_c}}}} \right)}_x}{{\left( {\frac{{\partial {T_{eff}}}}{{\partial x}}} \right)}_{{r_c}}} - {{\left( {\frac{{\partial V}}{{\partial x}}} \right)}_{{r_c}}}{{\left( {\frac{{\partial {T_{eff}}}}{{\partial {r_c}}}} \right)}_x}}}{{{{\left( {\frac{{\partial {P_{eff}}}}{{\partial x}}} \right)}_{{r_c}}}{{\left( {\frac{{\partial {T_{eff}}}}{{\partial {r_c}}}} \right)}_x} - {{\left( {\frac{{\partial {P_{eff}}}}{{\partial {r_c}}}} \right)}_x}{{\left( {\frac{{\partial {T_{eff}}}}{{\partial x}}} \right)}_{{r_c}}}}}\\
& = \frac{{48\pi r_c^2x(1 + {x^4})}}{{F(x,\omega )}}\left( {B'(x,\omega ) - \frac{{B(x,\omega )(3 + 7{x^4})}}{{x(1 + {x^4})}} - \frac{{B(x,\omega ){x^2}}}{{(1 - {x^3})}}} \right)
\end{aligned}
\end{equation}

where the differential of $D$ is given by $D'(x,\omega ) = \frac{{\partial D(x,\omega )}}{{\partial x}}$. It is important to further investigate the thermodynamic properties of a black hole at the critical point, besides the phase transition and critical behaviors. The critical point $x = {x_c}$ is determined at the position where the curves of $\alpha  - x$ and ${\kappa _{{T_{eff}}}} - x$ become infinitely divergent, which are explicitly illustrated in Fig. \ref{fig5}-\ref{fig6}. 

In term of the thermodynamic quantities discussed above, the extended Gibbs free energy for our system is given by
\begin{equation}  \label{3.20}
\begin{aligned}
G &= M - {T_{eff}}S\\
 &= {r_c}\left[ {\frac{{(1 - a)x(1 + x)}}{{2(1 + x + {x^2})}} + \frac{{{\phi ^2}(1 + x + {x^2} + {x^3})}}{{2x(1 + x + {x^2})}} - \frac{{{\beta _c}{x^3}}}{{6\pi \omega (1 - {x^3})}}(1 - {x^{ - 3 - 3\omega }}) - \frac{{B(x,\omega )(1 + {x^2} + f(x))}}{{4{x^3}(1 + {x^4})}}} \right]
\end{aligned}
\end{equation}

\begin{figure}[htp]
\centering
\begin{minipage}[t]{0.85\textwidth}
\subfigure[\quad $\phi=0.01, \beta _{c}=0.1, \omega=-1/2$]{\includegraphics[width=0.30\textwidth]{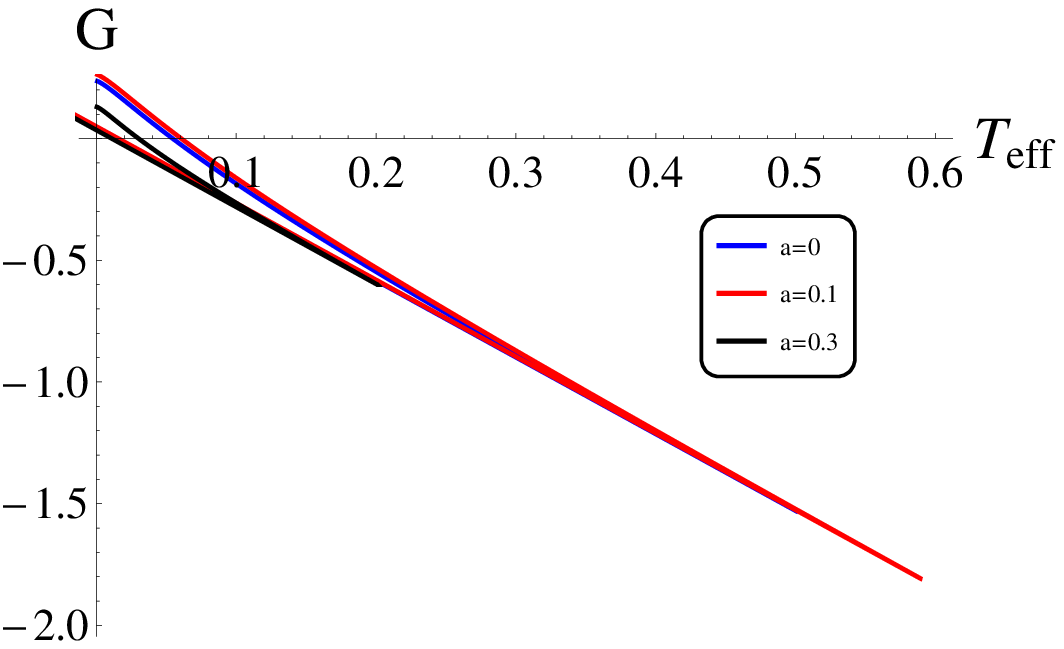}}
\subfigure[\quad$a=0.1, \phi=0.01, \beta _{c}=0.1$]{\includegraphics[width=0.30\textwidth]{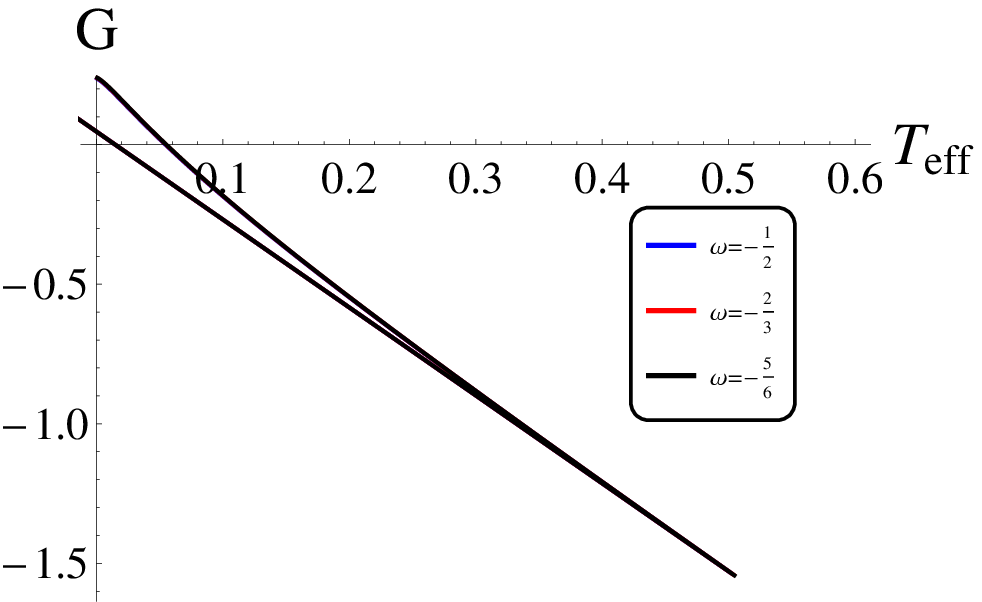}}\newline
\subfigure[\quad$a=0.1, \phi=0.01, \omega=- 1/2$]{\includegraphics[width=0.30\textwidth]{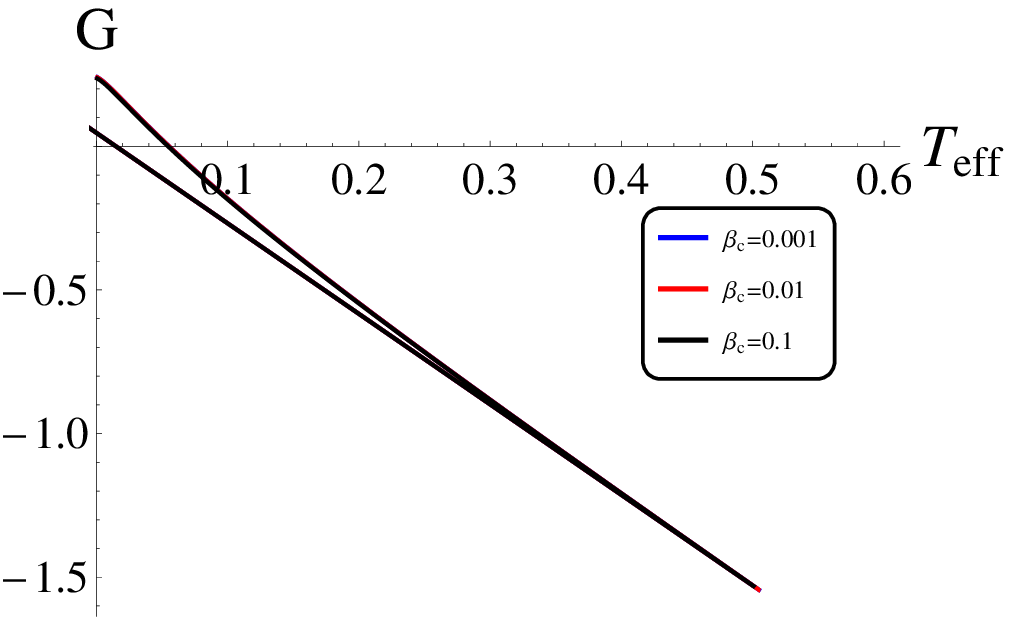}}
\subfigure[\quad$a=0.1, \beta _{c}=0.1, \omega=- 1/2$]{\includegraphics[width=0.30\textwidth]{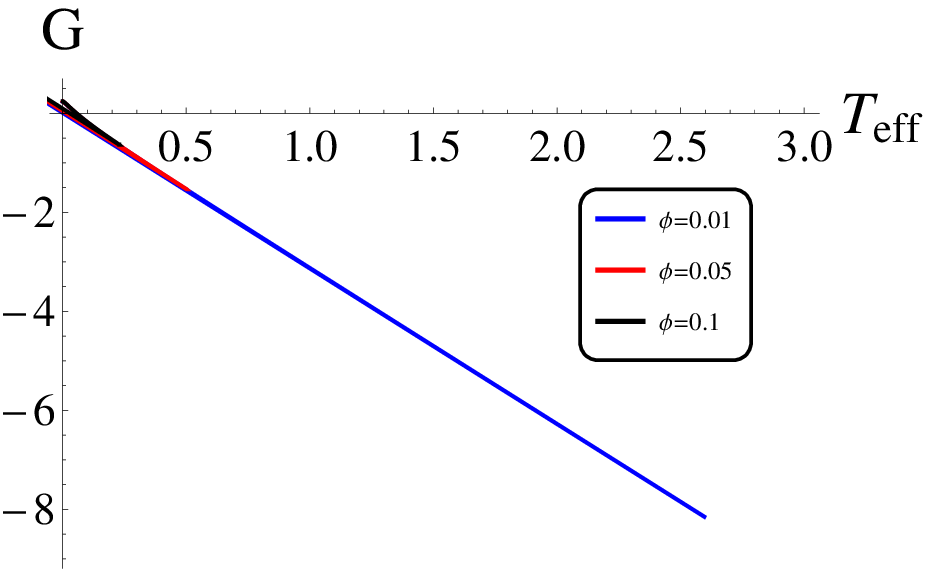}}\newline
\caption{The Gibbs function $G$ varying with effective temperature $T_{eff}$, concerning different combinations of relevant parameters ($a$, $\phi$, $\beta _{c}$, $\omega$).}
\label{fig7}
\end{minipage}
\end{figure}

Now, it is worthwhile commenting on the correlation between the corresponding Gibbs function and the effective temperature. Following the same procedure performed on the effective temperature, we obtain the $G - {T_{eff}}$ graphs presented in Fig. \ref{fig7}. Considering the fact that the entropy $S$ and the volume $V$ are both a continuous function of temperature $T_{eff}$, the critical point $x = {x_c}$ should also be a turning point in the $G - {T_{eff}}$ curve. In the following analysis of the thermodynamical study of critical phenomena, we will follow the detailed classification of general types of transition between phases of matter, firstly introduced by Paul Ehrenfest in 1933 and extensively applied in the previous literature \cite{Ma18}. More interestingly, in the framework of Ehrenfest¡¯s classification method, such point satisfies the requirements of a second-order phase transition, while the existence of first-order phase transitions can be ruled out in RN-dSSQ space-time. A quantitative analysis shown in Figs. \ref{fig4}-\ref{fig6} indicates that the criticality does not vary with the two parameters of $\beta _{c}$ and $\omega$, while the position of the critical point may dramatically change with different values of $a$ and $\phi$. Such behavior also resembles the behavior of effective temperature and effective pressure in RN-dSSQ space-time (Fig. \ref{fig1}-\ref{fig2}).

\section{Entropy force generated by the horizon interaction}

Motivated by the recent work by Ref.\cite{Verlinde11}, it was proposed that gravity can be explained as an entropic force caused by changes in the information associated with the positions of material bodies. In this scenario, the Newton¡¯s law of gravitation, the Einstein equations, and the law of inertia could be successfully derived from the entropic point of view. Concerning two independent ordinary thermodynamic systems, the entropy of the two systems are respectively $S_{A}$ and $S_{B}$. Therefore, in initial works it was always assumed that the entropy of the space-time is the sum of the entropy of the black hole horizon and the cosmological horizon. However, in this analysis, considering the interaction between the two space-time horizons, the entropy of RN-dSSQ space-time can be written as
\begin{equation}  \label{4.1}
S = {S_A} + {S_B} + {S_{AB}}
\end{equation}
where $S_{AB}$ denotes the external entropy generated by such interaction. In combination with the entropy definition in Eq. (\ref{3.3}), the total entropy of RN-dSSQ space-time can be divided into two parts, one part from the entropy of the black hole horizon and the cosmological horizon, and the other part from the contribution of the overall thermodynamic system, consist of the two thermodynamic systems corresponding to the two horizons. Moreover, when taking the mass $M$, the electric charge $Q$, and the space-time volume $V$ as the state parameters, the additional entropy generated by interactions between the two thermodynamic systems can be expressed as

\begin{equation}  \label{4.2}
{S_f} = \pi r_c^2f(x) = \pi r_c^2\left[ {\frac{8}{5}{{(1 - {x^3})}^{2/3}} - \frac{{2(4 - 5{x^3} - {x^5})}}{{5(1 - {x^3})}}} \right]
\end{equation}
From the entropic point of view, the effective entropic force due to the change in entropy obeys \cite{Verlinde11,Plastino18a,Plastino18b,Panos19,Kharzeev14,Cai10,Zhang17a,Zhang18a,Tahery00}
\begin{equation}  \label{4.3}
F =  - T\frac{{\partial S}}{{\partial r}}
\end{equation}
where $T$ is the temperature of the system, and $r$ is the radius location of the boundary surface. In our analysis, the entropic force due to interactions between the black hole horizon and the cosmological horizon of RN-dSSQ space-time is obtained as
\begin{equation}  \label{4.4}
F = {T_{eff}}{\left( {\frac{{\partial {S_f}}}{{\partial r}}} \right)_{{T_{eff}}}}
\end{equation}
with the effective temperature ${T_{eff}}$ and the sapec-time radius defined as $r = {r_c} - {r_ + } = {r_c}(1 - x)$. Combined with Eq. (\ref{4.2}), the above expression may be extended to
\begin{equation}  \label{4.5}
\begin{aligned}
F(x) &= {T_{eff}}\frac{{{{\left( {\frac{{\partial {S_f}}}{{\partial {r_c}}}} \right)}_x}{{\left( {\frac{{\partial {T_{eff}}}}{{\partial x}}} \right)}_{{r_c}}} - {{\left( {\frac{{\partial {S_f}}}{{\partial x}}} \right)}_{{r_c}}}{{\left( {\frac{{\partial {T_{eff}}}}{{\partial {r_c}}}} \right)}_x}}}{{(1 - x){{\left( {\frac{{\partial {T_{eff}}}}{{\partial x}}} \right)}_{{r_c}}} + {r_c}{{\left( {\frac{{\partial {T_{eff}}}}{{\partial {r_c}}}} \right)}_x}}}\\
&= \frac{{B(x)}}{{{x^3}(1 + {x^4})}} \times \frac{{2A(x)f(x) + x(1 + {x^4})B(x)f'(x)}}{{(1 - x)A(x) - (1 + {x^4})B(x)}}
\end{aligned}
\end{equation}
where
\begin{equation}  \label{4.6}
\begin{aligned}
A(x) = &{x^2}(1 - a)( - 1 - 3{x^2} + 6{x^3} - 5{x^4} - 4{x^5} + 9{x^6} - 6{x^7}) + {\phi ^2}(3 + 4{x^4} - 8{x^7} + 9{x^8})\\
&- \frac{{{\beta _c}{x^4}}}{{2\pi \omega }}\left[ {1 - 3{x^4} - 4\omega {x^3} - (1 - 3\omega ){x^{ - 3\omega }} - \omega (2 + 3\omega ){x^{ - 3 - 3\omega }} + 3(1 + \omega ){x^{4 - 3\omega }} - 3\omega (2 + \omega ){x^{1 - 3\omega }}} \right]\\
B(x) = &{x^2}(1 - a)(1 - 3{x^2} + 3{x^3} - {x^5}) - {\phi ^2}(1 - 3{x^3} + 3{x^4} - {x^7})- \frac{{{\beta _c}{x^4}}}{{2\pi \omega }}\left[ {1 - {x^{ - 3\omega }} - \omega ({x^3} - {x^{ - 3 - 3\omega }})} \right]
\end{aligned}
\end{equation}

\begin{figure}[htp]
\centering
\begin{minipage}[t]{0.85\textwidth}
\subfigure[\quad $\phi=0.01, \beta _{c}=0.1, \omega=-1/2$]{\includegraphics[width=0.31\textwidth]{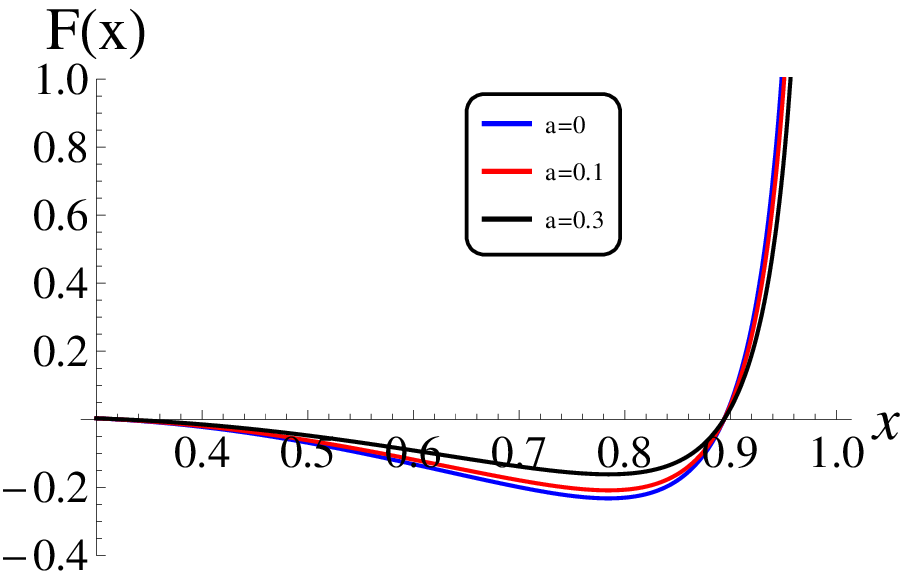}}
\subfigure[\quad$a=0.1, \phi=0.01, \beta _{c}=0.1$]{\includegraphics[width=0.30\textwidth]{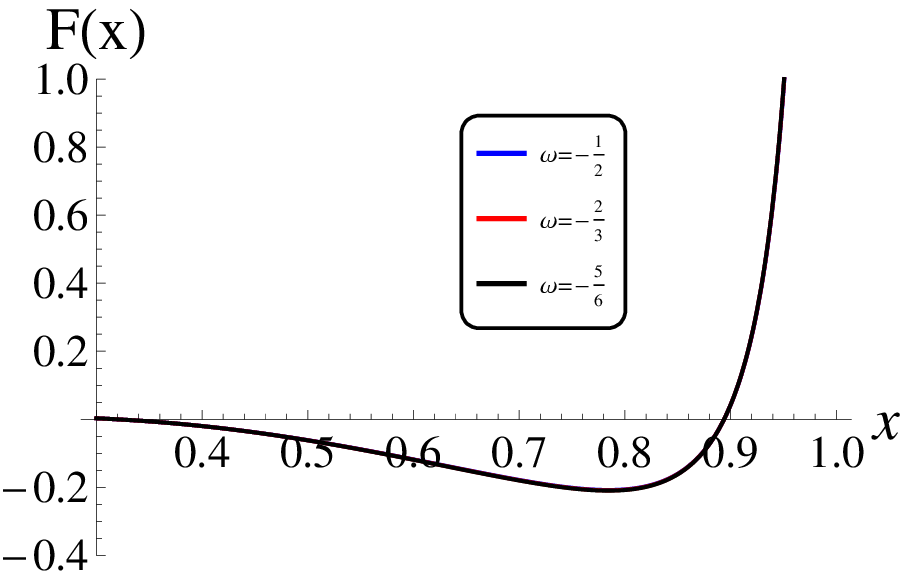}}\newline
\subfigure[\quad$a=0.1, \phi=0.01, \omega=- 1/2$]{\includegraphics[width=0.30\textwidth]{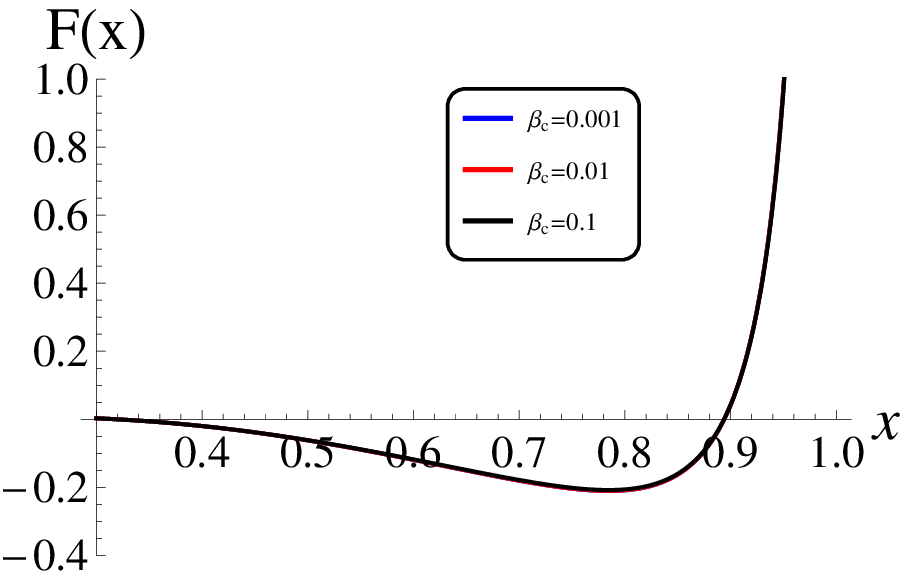}}
\subfigure[\quad$a=0.1, \beta _{c}=0.1, \omega=- 1/2$]{\includegraphics[width=0.30\textwidth]{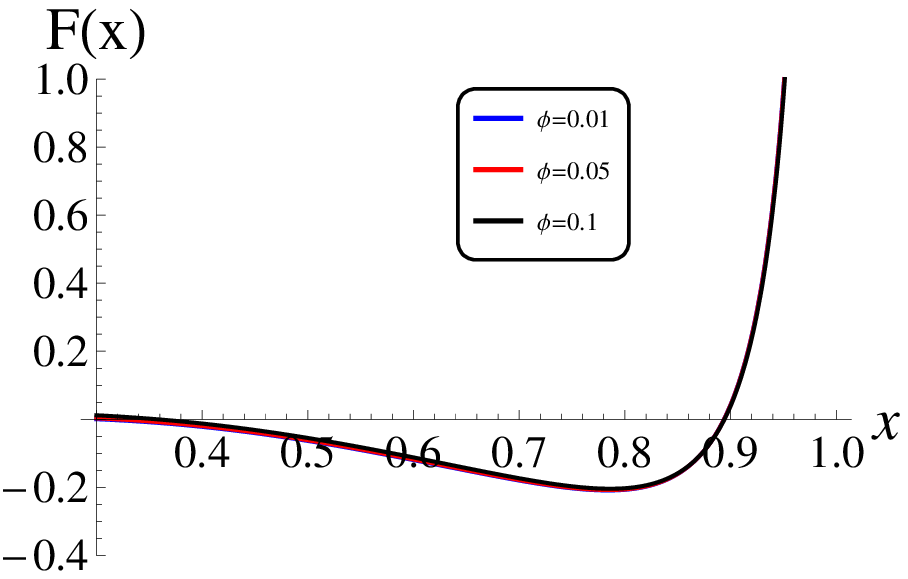}}\newline
\caption{The behavior of the entropy force as a function of $x$, (a) For different $a$, (b) for different $\omega$, (c) For different $\beta _{c}$, and (d) for different $\phi$.}
\label{fig8}
\end{minipage}
\end{figure}

In order to have a better illustration, the $F(x)-x$ diagram of the RN-dSSQ space-time is displayed in Fig. \ref{fig8} for the specific cases with different parameter combinations, which highlights the importance of the $a$ parameter. Therefore, compared with other effects quantified by three relevant parameters ($\phi$, $\beta _{c}$, $\omega$), the presence of the cloud of string plays significant role in determining the entropy force of RN-dSSQ space-time.

For a better comparison, in what follows we will concentrate on the interactions between neutral molecules or atoms with a center of mass separation $r$, which are often approximated by the so-called Lennard-Jones potential energy ${\phi _{L,J}}$ \cite{Plastino18a,Plastino18b,Miao18}

\begin{equation}  \label{4.7}
{\phi _{L,J}} = 4{\phi _{\min }}\left[ {{{\left( {\frac{{{r_0}}}{r}} \right)}^{12}} - {{\left( {\frac{{{r_0}}}{r}} \right)}^6}} \right]
\end{equation}

\begin{figure}
\begin{minipage}[htp]{0.485\linewidth}
\centering
\includegraphics[width=3in]{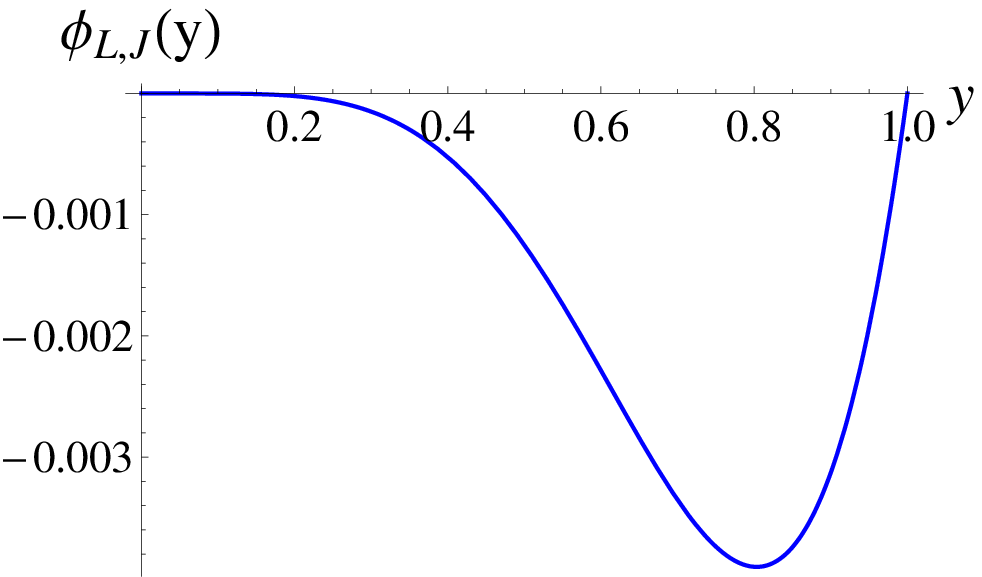}
\caption{The evolution of $\Phi_L - y$ with $\frac{{3 \times {2^{10}}{\phi _{\min }}}}{{{r_0}}} = 1$.}
\label{fig9}
\end{minipage}%
\begin{minipage}[htp]{0.485\linewidth}
\centering
\includegraphics[width=3in]{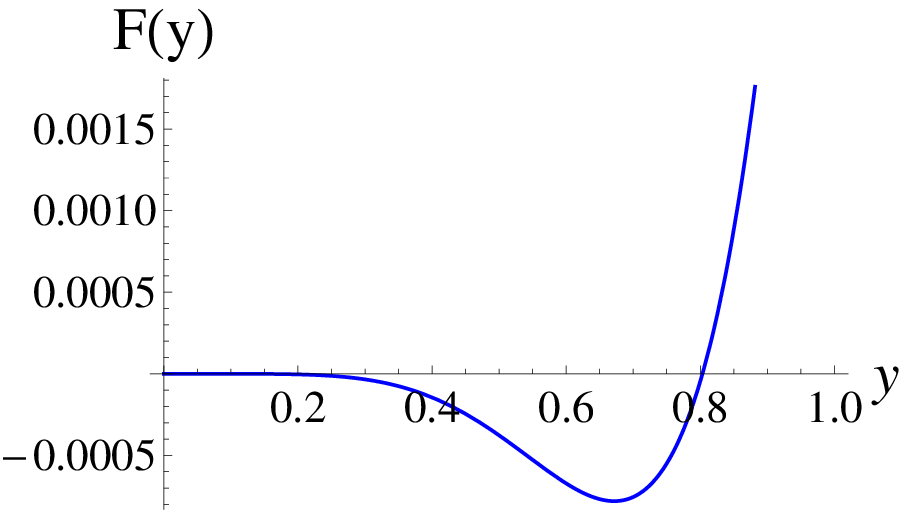}
\caption{The evolution of $F(y) - y$ with $\frac{{3 \times {2^{10}}{\phi _{\min }}}}{{{r_0}}} = 1$.}
\label{fig10}
\end{minipage}
\end{figure}

Note that the first term is a short-range repulsive interaction and the second term is a longer-range attractive interaction. A plot of  ${\phi _{L,J}}/{\phi _{\min }}$ versus $r/{r_0}$ is shown in Fig. \ref{fig9}. The value $r = {r_0}$ corresponds to ${\phi _{L,J}} = 0$, and the minimum value of ${\phi _{L,J}}$ is ${\phi _{L,J}}/{\phi _{\min }} =  - 1$ at

\begin{equation}  \label{4.8}
{r_{\min }}/{r_0} = {2^{1/6}} \approx 1.122
\end{equation}

Based on the definition of potential energy, the force between a molecule and a neighbor in the radial direction from the first molecule is ${F_r} =  - d{\phi _{L,J}}/dr$, which is positive (repulsive) for $r < {r_{\min }}$ and negative (attractive) for $r > {r_{\min }}$. When the center of the first atom (with the radius of ${r_0}/2$) is well coincided with the coordinate center, the corresponding center of the second atom (with the boundary of ${r_2}$) is $r={r_2}+ {r_0}/2$. When taking $y={r_0}/(2{r_2})$, and $0 < y \le 1$, one may transform Eq.(\ref{4.7}) to obtain the expression of Lennard-Jones potential energy
\begin{equation}  \label{4.9}
{\phi _{L,J}}(y) = 4{\phi _{\min }}\left[ {{{\left( {\frac{{{r_0}}}{r}} \right)}^{12}} - {{\left( {\frac{{{r_0}}}{r}} \right)}^6}} \right] = 4{\phi _{\min }}{2^6}\left[ {{2^6}{{\left( {\frac{y}{{1 + y}}} \right)}^{12}} - {{\left( {\frac{y}{{1 + y}}} \right)}^6}} \right]
\end{equation}
the evolution of which is displayed in Fig. \ref{fig9}. Furthermore, it is straightforward to obtain the interaction between the two atoms as
\begin{equation}  \label{4.10}
F(y) =  - \frac{{d{\phi _{L,J}}}}{{dr}} = 4{\phi _{\min }}\frac{6}{r}\left[ {2{{\left( {\frac{{{r_0}}}{r}} \right)}^{12}} - {{\left( {\frac{{{r_0}}}{r}} \right)}^6}} \right] = \frac{{3{\phi _{\min }}{2^{10}}}}{{{r_0}}}\left[ {{2^7}{{\left( {\frac{y}{{1 + y}}} \right)}^{13}} - {{\left( {\frac{y}{{1 + y}}} \right)}^7}} \right]
\end{equation}

Now, it is worthwhile commenting on the correlation between the entropy force generated by the interaction between the two horizons and the Lennard-Jones force between two particles. Our results demonstrate the strong degeneracy between the entropy force of the two horizons and the ratio of the horizon positions, which follows the surprisingly similar law given the relation between the Lennard-Jones force and the ratio of two particle positions. More specifically, in order to obtain a better comparison between the two types of forces, we put the two curves (Fig. \ref{fig8} and Fig. \ref{fig10}) in the same coordinates.
\begin{figure}[htp]
\begin{minipage}[t]{0.85\textwidth}
   \centering
   \includegraphics[width=4in]{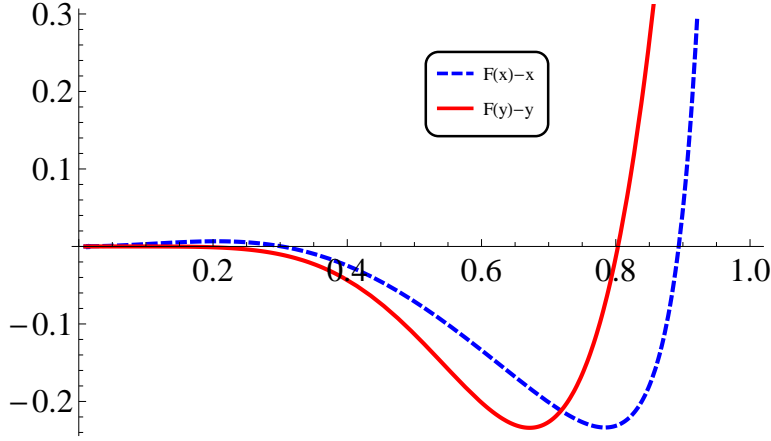}
    \caption{The comparison between the $F(x) - x$ graph with $a = 0$, $\phi = 0.01$, ${\beta _c} = 0.1$, $\omega  =  - 2/3$ (dashed line) and the $F(y) - y$ graph with $\frac{{3 \times {2^{10}}{\phi _{\min }}}}{{{r_0}}} = 300$ (solid line).}
\label{fig11}
\end{minipage}
\end{figure}
The results shown in Fig. \ref{fig11}, again, confirm the above conclusion. In particular, when $x$ approaches one $x \to 1$, the entropy force of the two horizons reaches its maximum value and thus generates the cosmic acceleration; when the ratio of the horizon positions reaches to a certain value, the entropy force of the two horizons becomes negative and then gravity should lead to a slowing of the expansion; at the critical point $x = {x_c}$, the Universe will evolve into a new stage of de Sitter phase and exhibits the slowing expansion due to gravity; finally, when $x$ approaches zero $x \to 0$, the entropy force of the two horizons reaches zero and thus generates constant expansion rate. Therefore, the exploration of the entropy force between the black hole horizon and the cosmological horizon will not only contribute to deep understanding the evolution of our Universe (inflation in the early universe and cosmic acceleration in the late time), but also provide a new window to differentiate the three typical modes of cosmic expansion.

\section{Conclusion and discussion}

\label{sec:conclusion}
It is well known that black hole is an ideal system to understand the nature and behavior of quantum gravity. More importantly, black hole thermodynamics continues to be one of great importance in gravitational physics. In this paper, in the framework of a Reissner-Nordstr\"om-de Sitter black hole, we discuss the combined effects of the cloud of strings and quintessence on its thermodynamic properties, by considering the interaction between the black hole horizon and the cosmological horizon. Here we summarize our main conclusions in more detail:

1. Focusing on the derivation of the equivalent thermodynamic quantities, we obtain the effective temperature, the effective pressure, and the total entropy of the RN-dSSQ space-time, and furthermore quantify the effects of relevant state parameters ($a$, $\phi$, $\beta _{c}$, $\omega$). One can clearly see that the maximum value of the effective temperature will significantly decrease with the value of $a$ and $\phi$, which is quite similar to the behavior of effective pressure.

2. The effective isovolumetric heat capacity of RN-dSSQ space-time is not equal to zero, which is quite different from with the case of in AdS space-time. However, such finding in RN-dSSQ space-time agrees very well with the previous result obtained in the framework of ordinary thermodynamic systems, which strongly implies the strong consistency between the two types of thermodynamic systems.

3. In the framework of RN-dSSQ space-time, our results show that the entropy S is an explicit function of the horizon position (the black hole horizon and the cosmological horizon), instead of the function of the electric charge $Q$ and cosmological constant $l^2$. Meanwhile, the function $f(x)$ , which represents the extra contribution from the correlations of the two horizons, is also well consistent with the form in the RN-dS space-time.

4. In the analysis of the thermodynamic study of critical phenomena, our findings prove that similar to the case in AdS space-time, second-order phase transitions could take place under certain conditions, with the absence of first-order phase transition in the charged de Sitter black holes with cloud of string and quintessence. More importantly, a quantitative analysis indicates that the criticality does not vary with the two parameters of $\beta _{c}$ and $\omega$, while the position of the critical point may dramatically change with different values of a and $\phi$. Therefore, our results could provide a new approach to study the thermodynamic qualities of unstable dS space-time, as well as a theoretical foundation for investigating the phase transition and phase structure of RN-dSSQ black hole systems.

5. Working on the entropy force generated by the interaction between the black hole horizon and the cosmological horizon, we find that the relation between the entropy force of the two horizons and the ratio of the horizon positions follows the similar law between the Lennard-Jones force and the ratio of two particle positions. Therefore, the exploration of the entropy force between the black hole horizon and the cosmological horizon will not only contribute to the construction of the overall evolution history of the Universe (three typical modes of cosmic expansion), but also shed light on the possible interactions between the particles in black holes.

\section*{Acknowledgments}

We thank Prof. Z. H. Zhu for useful discussions.

This work was supported by the National Natural Science Foundation of China (Grant Nos. 11847123, 11475108, 11705106, 11705107, 11605107), the Natural Science Foundation of Shanxi Province, China (Grant No. 201601D102004). The
Initial Foundation of Mianyang Teachers' College (Grant No. QD 2016A002), Natural Science Foundation of Education Department in Sichuan Province (Grant No. 17ZB0210).

\end{document}